\documentclass{aastex62}
\usepackage[version=4]{mhchem}
\usepackage[normalem]{ulem}
\usepackage{CJK}
\graphicspath{{./}{figures/}}

\received{xxx, 2019}
\revised{xxx, 2019}
\accepted{\today}
\submitjournal{ApJ}

\shorttitle{3D projection effects on chemistry}
\shortauthors{Ge et al.}

\begin{document}

\title{Three dimensional projection effects on chemistry in a Planck galactic cold clump}

\correspondingauthor{J. X. Ge}
\email{jge@das.uchile.cl;gejixing666@gmail.com}

\author[0000-0002-9600-1846]{J. X. Ge}
\affil{Departamento de Astronom\'{\i}a, Universidad de Chile, Camino el Observatorio 1515, Las Condes, Santiago, Chile}

\author{Diego Mardones}
\affil{Departamento de Astronom\'{\i}a, Universidad de Chile, Camino el Observatorio 1515, Las Condes, Santiago, Chile}
\affiliation{Centre for Astrochemical Studies, Max-Planck-Institute for Extraterrestrial Physics, Giessenbachstrasse 1, 85748, Garching, Germany}
	
\author[0000-0002-3938-4393]{J. H. He}
\affil{Yunnan Observatories, Chinese Academy of Sciences, 396 Yangfangwang, Guandu District, Kunming, 650216, P. R. China}
\affiliation{Chinese Academy of Sciences South America Center for Astronomy, National Astronomical Observatories, Chinese Academy of Sciences, Beijing 100012, People's Republic of China}
\affil{Departamento de Astronom\'{\i}a, Universidad de Chile, Camino el Observatorio 1515, Las Condes, Santiago, Chile}

\author{Jonathan M C Rawlings}
\affiliation{Department of Physics and Astronomy, University College London, Gower Street, London WC1E 6BT, UK.}

\author{Sheng-Yuan Liu}
\affiliation{Institute of Astronomy and Astrophysics, Academia Sinica. 11F of Astronomy-Mathematics Building, AS/NTU No. 1, Section 4, Roosevelt Rd., Taipei 10617, Taiwan}

\author{Jeong-Eun Lee}
\affiliation{School of Space Research, Kyung Hee University, Yongin-si, Korea}

\author{Ken'ichi Tatematsu}
\affiliation{National Astronomical Observatory of Japan, National Institutes of Natural Sciences, 2-21-1 Osawa, Mitaka, Tokyo 181-8588, Japan}
\affiliation{Department of Astronomical Science, SOKENDAI (The Graduate University for Advanced Studies), 2-21-1 Osawa, Mitaka, Tokyo 181-8588, Japan}

\author{Tie Liu}
\affiliation{Shanghai Astronomical Observatory, Chinese Academy of Sciences, 80 Nandan Road, Shanghai 200030, China}
\affiliation{East Asian Observatory, 660 N. A’ohoku Place, Hilo, HI 96720, USA}

\author{Lei Zhu}
\affiliation{Chinese Academy of Sciences South America Center for Astronomy, National Astronomical Observatories, Chinese Academy of Sciences, Beijing 100012, People's Republic of China}

\author{Qiang Chang}
\affiliation{School of Physics and Optoelectronic Engineering, Shandong
	University of Technology, Zibo 255000}

\author{Natalia Inostroza}
\affil{Universidad Aut{\'o}noma de Chile, Facultad de Ingenier{\'i}a, N{\'u}cleo de Astroqu{\'i}mica \& Astrof{\'i}sica, Av. Pedro de Valdivia 425,Providencia, Santiago, Chile.}

\author[0000-0002-4707-8409]{S. Feng}
\affiliation{National Astronomical Observatory of China, Datun Road 20, Chaoyang, Beijing, 100012, P. R. China}
\affiliation{CAS Key Laboratory of FAST, NAOC, Chinese Academy of Sciences}
\affiliation{National Astronomical Observatory of Japan, National Institutes of Natural Sciences, 2-21-1 Osawa, Mitaka, Tokyo 181-8588, Japan}



\begin{abstract}

Offsets of molecular line emission peaks from continuum peaks are very common but frequently difficult to explain with a single spherical cloud chemical model. We propose that the spatial projection effects of an irregular three dimensional (3D) cloud structure can be a solution. This work shows that the idea can be successfully applied to the Planck cold clump G224.4-0.6 by approximating it with four individual spherically symmetric cloud cores whose chemical patterns overlap with each other to produce observable line maps. With the empirical physical structures inferred from the observation data of this clump and a gas-grain chemical model, the four cores can satisfactorily reproduce its 850\,$\mu$m continuum map and the diverse peak offsets of CCS, \ce{HC3N} and \ce{N2H+} simultaneously at chemical ages of about $8\times 10^5\sim 3\times 10^6$\,yrs. The 3D projection effects on chemistry has the potential to explain such asymmetrical distributions of chemicals in many other molecular clouds.

\end{abstract}


\section{Introduction}
Molecular lines are powerful tools to diagnose the physical and chemical properties of molecular clouds, but the spatial distributions in line maps are frequently found to deviate from the \ce{H2} column density distributions, which is not easy to interpret with a single point chemical model or one based on a single spherical cloud core model. In such cases, the emission peaks of the molecular lines are definitely offset from and strongly asymmetric with respect to the \ce{H2} column density peaks or ridges in the clouds. 
For example, \citet{Spezz2017} found in the well known dense core L1544 three families of molecular lines, namely the HNCO, \ce{CH3OH} and \ce{c-C3H2} families, whose peak positions significantly deviate from the continuum peak. There is no convincing interpretation yet.

Such phenomena have also been found in our large single dish survey campaign TOP-SCOPE \citep{Liutie2018} toward a large sample of Planck Galactic Cold Clumps (PGCCs) \citep{Planck2011c,Planck2016}. The first TOP-SCOPE line observation paper by \citet{Tate2017} revealed large asymmetric offsets of CCS and \ce{HC3N} line emission peaks from 850\,$\mu$m continuum peaks in several sources. The target of this work, G224.4-0.6, is just the most prominent example among them. Both CCS and \ce{HC3N} lines show a single emission peak that offset to one side of the continuum peaks and even the emission peak of the dark cloud tracer line of \ce{N2H+} also shows a smaller offset of about one telescope beam. 
The former two molecules are expected to show depletion effects in dark clouds \citep[e.g.,][]{Aikawa2001,Shinnaga2004}. Although the last one is expected to peak at the center of cold cores after the depletion of CO \citep[see e.g.,][]{Bergin1997,Benson1998,Caselli1999,Bergin2001,Caselli2002}, depletion will still occur when the chemical age is long enough to allow depletion of its parent species \ce{N2} \citep[e.g.][]{Bergin2002}; this has been observed in starless cores \citep{Lippok2013,Redaelli2019}.  
However, the observed offset line-emission peaks can not be interpreted by such a chemical picture in a single round or elongated core.

In this paper, we introduce the idea of overlapping of chemical patterns of multiple independent cores due to 3D projection effects to explain the above phenomena. After a more detailed description of the chemical problem in our target source G224.4-0.6 in Section~\ref{sec_chem_pro}, we constrain the physical models of four cloud cores from a continuum map (Section~\ref{sec_phy}) and construct gas-grain chemical models for each of them (Section~\ref{sec_chem}). In Section~\ref{sec_comp}, we empirically adjust the physical model parameters of some cores to find a solution that well reproduce both the continuum and the line emission maps of CCS, \ce{HC3N} and \ce{N2H+} simultaneously. The best chemical model and the proposed chemical overlapping effect is briefly discussed in Section~\ref{sec_discussions}. Finally, Section~\ref{sec_con} summarizes our main findings.

\section{Observations and chemical problems in G224.4-0.6}
\label{sec_chem_pro}
\subsection{Observations}
\begin{figure*}
	\centering
	\includegraphics[scale=0.72]{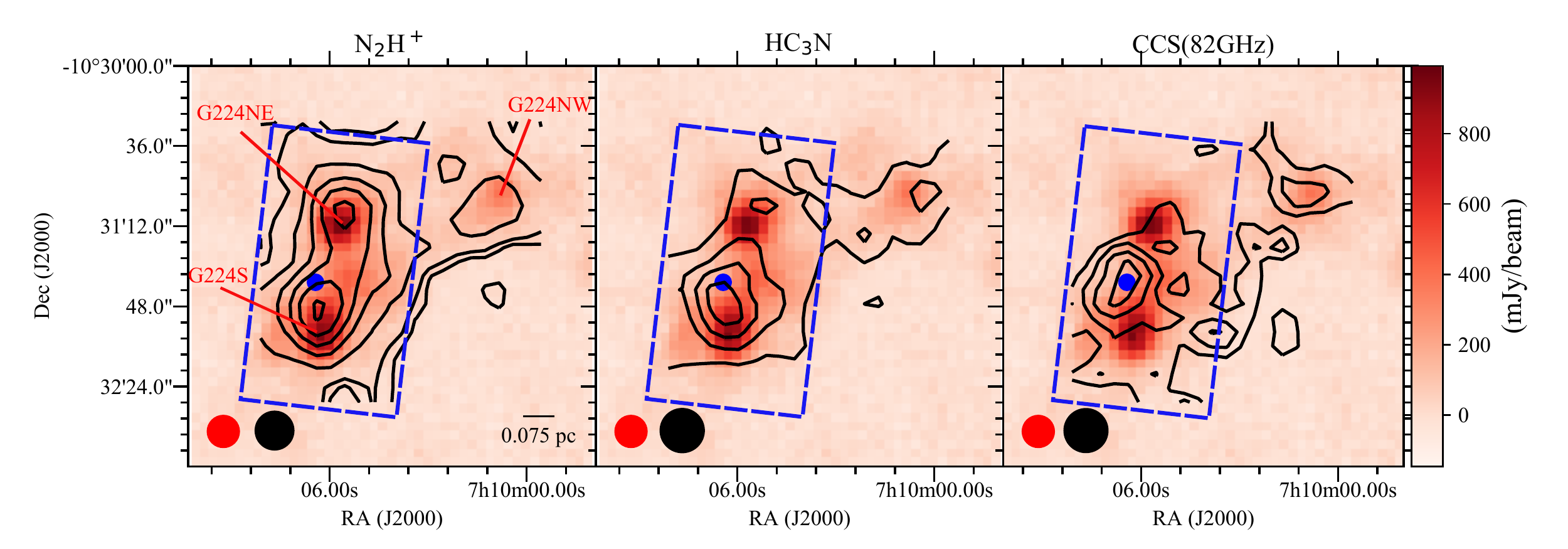}
	\caption{The 850\,$\mu$m continuum image (color scale) of G224.4-0.6 overlaid with the integrated intensity maps (contours) of three molecular lines: the \ce{N2H+} line ({\em left}) at levels of [3, 5, 10, 15, 20, 25, 30]\,$\times 0.15$\,K\,km\,s$^{-1}$ ($1\sigma$), the \ce{HC3N} line ({\em middle}) at levels of [3, 6, 9, 12]\,$\times 0.12$\,K\,km\,s$^{-1}$ ($1\sigma$) and the 82\,GHz CCS ({\em right}) at levels of [3, 4.5, 6, 7.5]\,$\times 0.13$\,K\,km\,s$^{-1}$ ($1\sigma$). The corresponding half-power beam sizes are shown with corresponding colors at the bottom left corner. The blue point marks the CCS peak position. The blue rectangle indicates the region used to constrain the physical and chemical models in this work.}
	\label{fig_G224_obs}
\end{figure*}

G224.4-0.6 is a Planck cold clump with dense cores \citep{Planck2011c,Planck2016}. It was selected from ECC (early cold core) catalog with the PLCKECC (Planck early cold cores) name G224.47-00.65 by \citep{Liutie2018}. The distance to the clump is unknown. But it is less than $1^\circ$ from the massive cloud complex CMa OB1 region for which \citep{Kim1996} assigned a distance of $\sim 1100$\,pc from the study of the OB stars in this region by \citet{Claria1974}. We adopt this distance in the current study. We have mapped G224.4-0.6 in 850\,$\mu$m continuum using JCMT/SCUBA-2 and three 3\,mm molecular lines, \ce{N2H+}\,$J=1-0$, \ce{HC3N}\,$J=9-8$, and CCS $J_N=7_6-6_5$ (82\,GHz), with the Nobeyama 45\,m radio telescopes \citep{Tate2017}. The results are shown in Fig.~\ref{fig_G224_obs} (continuum in color scale and lines in contours). The beam size of the SCUBA-2 continuum map is $14''$, which comparable to the $18.8\pm0.3''$ beam of the line maps. The region has two major dust emission peaks near the center of the maps, together with another fainter core to the north west. This chemistry study concentrates on the sub-region around the two cold cores represented by the two major peaks, which were named as G224S (south) and G224NE (north-east) by \citet{Tate2017} respectively. The spectral resolution was about $0.05\sim 0.06$\,km\,s$^{-1}$.

\subsection{Chemical problems}
From Fig.~\ref{fig_G224_obs}, the \ce{N2H+} line shows two peaks that coincide with the continuum peaks well, except a very small offset from the G224S peak. The \ce{HC3N} line mainly possesses a strong core near to G224S but with a bit larger position offset toward the northeast direction, together with another much fainter core near to G224NE. The most interesting case is the CCS line which mainly shows a single peak that coincide with neither the continuum peaks nor the line connecting the two peaks. The CCS peak position is also marked by a blue dot in all three panels, which demonstrates that it does not agree with any peaks of the other two molecules as well. 
Although the lower abundance of CCS toward the two dust continuum peaks could be explained by the effect of depletion in the core centers \citep[e.g.][]{Aikawa2001,Shinnaga2004}, the offset of the CCS peak to one side of the clump is not understood. We will try to explain this with the coupling of asymmetrical cloud structures and dark cloud gas grain chemistry in this work.

\section{Physical and chemical model of multiple cores}\label{sec_phy}

\begin{figure}
	\centering
	\includegraphics[scale=0.8]{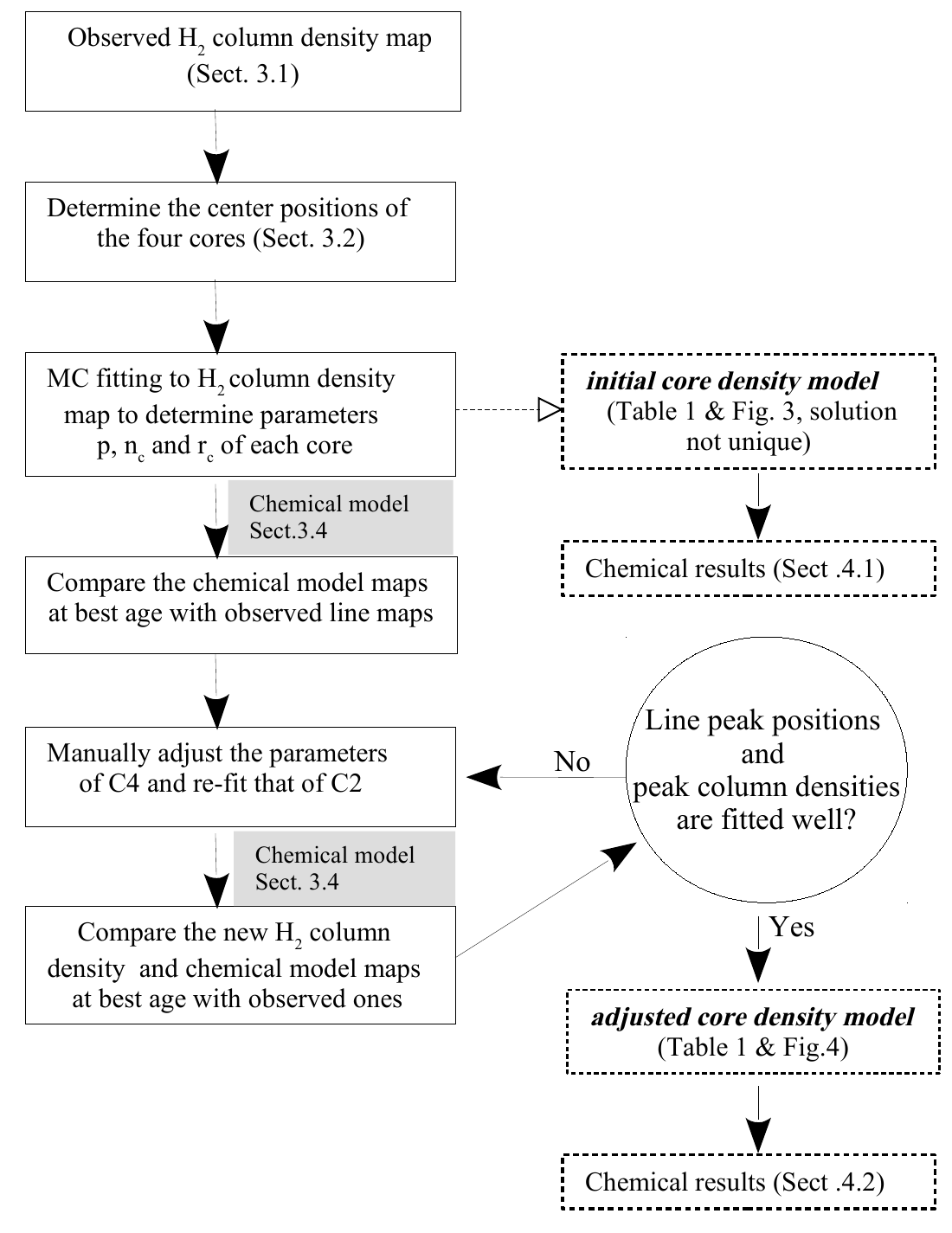}
	\caption{Work flow to determine the physical model of the cores. MC = Monte Carlo.
	}
	\label{fig:workflow}
\end{figure}

In order to construct the physical and chemical models of the cloud, we will start from fitting the azimuthally averaged radial gas column density profiles of the two major cores G224NE and G224S (this part is put in Appendix~\ref{sec_fits_1d}) and then extend it to directly fitting a four-core model to the observed 2-D maps. From the fitting to the observed \ce{H2} map alone, a {\em initial core density model} will be defined. Then the core parameters will be manually modified to improve the chemical behavior of the model and finally an adjusted core density model will be adopted. The detailed procedures are given in the subsections below and the whole procedure is summarized by the flow chart in Fig~\ref{fig:workflow}.

\subsection{\ce{H2} column density map}\label{subsec:NH2}
To obtain the physical structure of G224.4-0.6, we first convert the SCUBA-2 850\,$\mu$m dust continuum map to H$_2$ column density map using the formula proposed by \citet{Kauffmann2008}:
\begin{equation}
\label{eq_NH2}
\begin{split}
N({\rm H_2})=
& 2.02\times 10^{20}\,{\rm cm}^{-2}  \left( {\rm e}^{1.439(\lambda/{\rm mm})^{-1}(T_{\rm d}/10\,{\rm K})^{-1}} -1 \right) \\
& \left( \frac{\kappa_{\nu}}{\rm 0.01\,cm^2\,g^{-1}}\right)^{-1} \left( \frac{S_{\nu}^{beam}}{\rm mJy\,beam^{-1}}\right) \\
& \left( \frac{\theta_{\rm HPBW}}{\rm 10\,arcsec}\right)^{-2} \left( \frac{\lambda}{\rm mm}\right)^3
\end{split}
\end{equation}
where the dust temperature $T_{\rm d}$ is set to 11.1\,K according to \citet{Planck2016}. $S_{\nu}^{\rm beam}$ is the observed flux per beam with a half power beam width of $\theta_{\rm HPBW}$, and $\kappa_{\nu}=0.0182$\,cm$^2$\,g$^{-1}$ is the dust opacity at 850\,$\mu$m for dust models with thin dust mantles \citep{Ossenkopf1994}. This value was also widely used to estimate column density and mass \citep[e.g.][]{Kauffmann2008}. Here the dust-to-gas mass ratio of 0.01 is used.

The derived \ce{H2} column density map used in this work is shown in the left panel of Fig.~\ref{fig_fits_2d} (whole map can be found in left panel of Fig.~\ref{fig_fits_C1C2} in Appendix~\ref{sec_fits_1d}). Beside the three major cores whose highest peak column densities reach $\sim 1.2\times 10^{23}$\,cm$^{-2}$, some fainter extended structures are also visible around G224S, which will be essential to solve the chemistry problem in later sections. Due to the nature of SCUBA-2 observation, there are some pixels with negative flux in Fig.~\ref{fig_G224_obs} (in pale color) which prevent us from estimating the column densities. Therefore, we can adopt the background gas column density of $\sim 4\times 10^{21}$\,cm$^{-2}$ from Herschel data \citep{Marsh2017} for these faint regions when necessary, though the Herschel spatial resolution at sub-mm is slightly lower than the SCUBA-2 data.

\subsection{Core density structures}\label{sec_fits_2d}

Because the cloud cores are found to be small and the SCUBA-2 beam did not resolve them well, it is challenging to find a unique density structure model for each of them. We have performed trial fittings to the azimuthally averaged column density profiles of the two main cores G224NE and G224S with 1-D density models (Appendix~\ref{sec_fits_1d}) and perfectly illustrated the strong parameter degeneracy for each core. Our purpose in this section is not to determine unique solutions of the density structures of the cores from the observed column density map, but to find a reasonably good physical model that would allow us to explore possible ways to reproduce the observed molecular chemistry features. 

\begin{figure}
	\centering
	\includegraphics[width=0.55\linewidth]{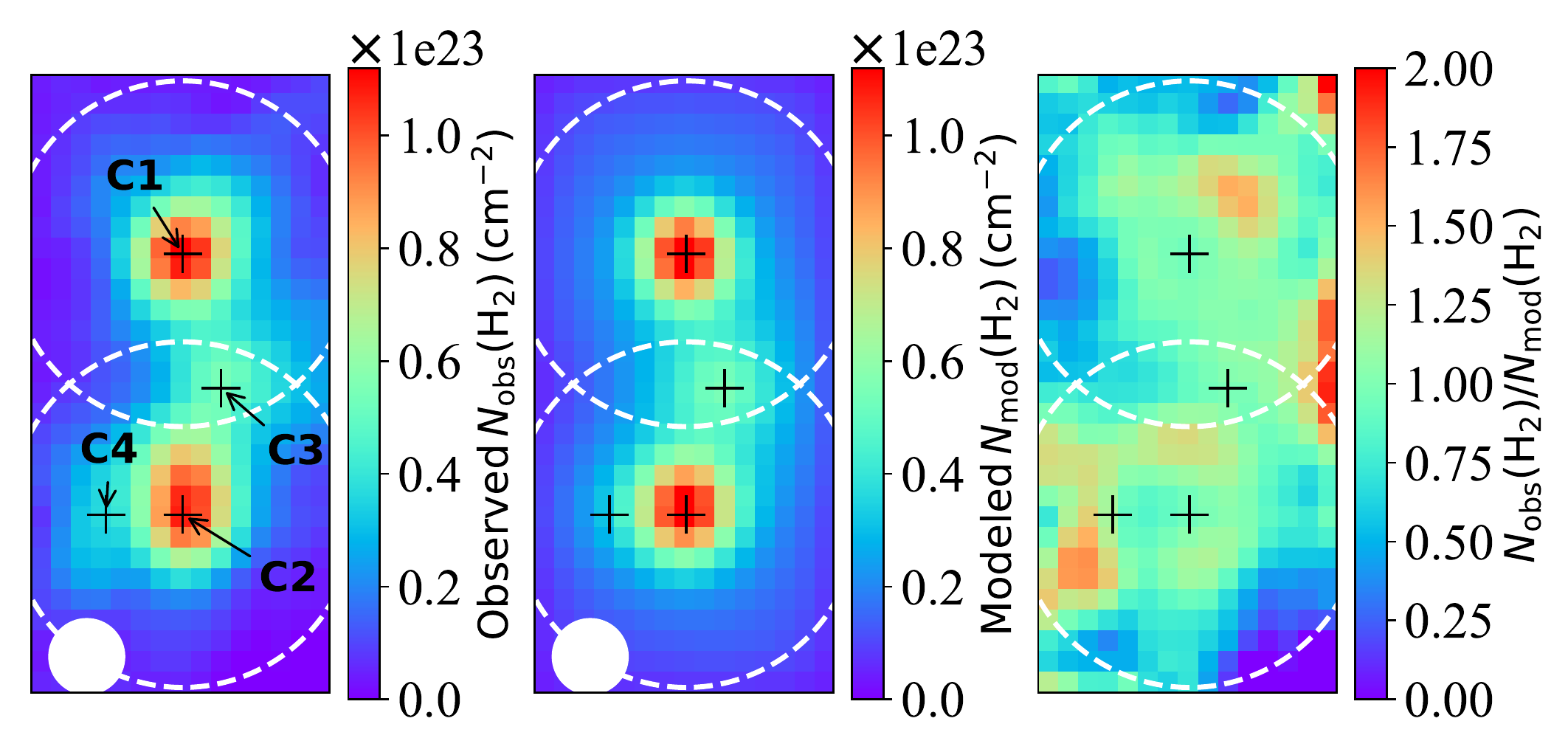}
	\caption{Comparison between observed (left) and modeled (middle) \ce{H2} column density maps. The right panel shows the ratio between observed and modeled maps. The black crosses indicate the centers of the four cloud cores. The large dashed-line circles show the outer radius $R$ of the two major cores C1 and C2. Images in all panels of this figure and Fig.~\ref{fig_reasonable_phy} have been rotated by 8 degrees anticlockwise to facilitate comparison by eyes, with each one corresponding to the blue rectangle region in Fig.~\ref{fig_G224_obs}.}
	\label{fig_fits_2d}

	\includegraphics[width=0.55\linewidth]{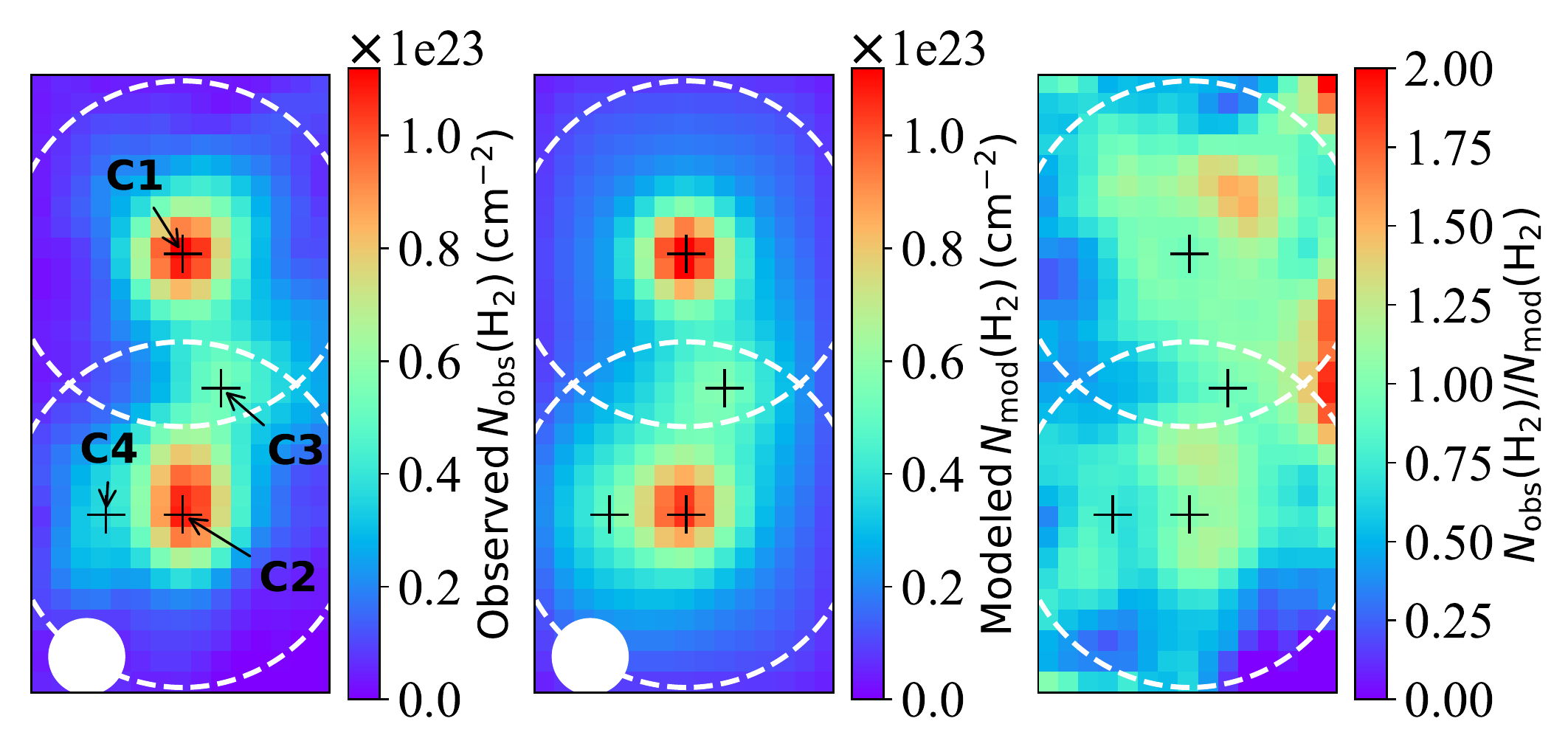}
	\caption{Same as Fig~\ref{fig_fits_2d}, but with improved parameters of C2 and C4. See the details in the text.}
	\label{fig_reasonable_phy}
\end{figure}

The most salient chemical features in this source is the offsets of the CCS and \ce{HC3N} emission peaks from the symmetry axis of the two major cores. Hereafter, we call G224NE as C1 and G224S as C2 and chose their core center positions to be the observed continuum flux peak pixels. Thus, solely the two cores are insufficient to explain the asymmetrical chemical feature. In the column density map in the left panel of Fig.~\ref{fig_fits_2d}, a fainter core appears between C1 and C2 that can be named C3. The core center can be directly determined as the column density peak there. Its position is offset by about 7\,arcsec to the west of the C1-C2 symmetry axis. Because both CCS and \ce{HC3N} are expected to be depleted in cold core center, their abundance peaks will appear in a shell structure (or a ring-like structure in the sky plane). We expect the overlapping of the molecular abundance rings of the core C3 with that of C1 and C2 would help break the C1-C2 symmetry in the molecular line emission maps. 

However, after some experiments of the chemical simulations with the three cores, as we will show in later sections, we recognize that another core C4 to the east side of C2 is essential to explaining the observed peak positions in both the CCS and \ce{HC3N} line maps simultaneously. The observed column density map in the left panel of Fig.~\ref{fig_fits_2d} indeed shows an faint extended structure to the east side of C2. We pick up a position near the center of this extended structure, offset to the east of C2 by 14\,arcsec, as the core center for C4. But more chemical modeling experiments have shown that the exact position of C4 is not a key factor. The center positions of all four core are shown in Table~\ref{tab_fits_2d}.

For each cloud core, we adopted the Plummer-like function \citep{Plummer1911} to simplify our model fitting problem because we do not concentrate on the differentiation of different kinds of density models (e.g the Bonnor-Ebert density distribution) due to too large telescope beams of our data. The plummer-like 1-D core density is expressed as
\begin{equation}
\label{eq_nH}
n(r)=\frac{n_c}{[1+(r/r_c)^2]^{p/2}} + n_{\rm b}
\end{equation}
where $n_c$ is the core-center density, $r_c$ is the radius of the central flat inner core, $p$ is the power-low exponent at large radius ($r\geq r_c$) and $n_{\rm b}$ is the background gas density. Generally, the $p=4$ case corresponds to an isothermal gas cylinder in hydrostatic equilibrium \citep{Ostriker1964}, while the $p=2$ case mimics magnetohydrodynamic models \citep[e.g.][]{Fiege2000}.

We try to find a reasonable density model for each core by fitting the 2-D column density model maps of the four cores to the SCUBA-2 column density map. To reduce the size of parameter space, we first fix the background gas density $n_{\rm b}$ and the outer radius $R$ of each core. For the major cores C1 and C2, $n_{\rm b}$ is set to the value of $10^4$\,cm$^{-3}$ found from our 1-D column density profile fitting in Appendix~\ref{sec_fits_1d}, while that of the two fainter cores is set to zero. The choices of the different $n_{\rm b}$ values are based on the fact that C1 and C2 are more massive than C3 and C4 so that the former two cores could be more extended than observed and modeled in this work. The outer radius $R$ is set to the same value of 0.17\,pc for all four cores; this value is constrained by avoiding the appearance of negative pixels for the two major cores C1 and C2 in the observed column density map. Then, the free parameters of each cloud core will be only $p$, $n_c$, and $r_c$. The total number of free parameters for the four cores is 12. However, similarly due to the too low spatial resolution of our maps, as we mentioned in Appendix~\ref{sec_fits_1d}, there still exists strong degeneracy among the three parameters for each core.

For any set of given parameters of a cloud core, we construct its column density model by first producing a 3-D spherical density distribution according to the 1-D formula in Eq.~\ref{eq_nH} and projecting it onto the sky plane to obtain the column density map. Then, the four cores are put to the corresponding observed core-center positions and merged together to form the 2-D model column density map. After being convolved by the SCUBA-2 beam of $14''$, it can be compared with the observed 2-D column density map to define the posterior likelihood value used in the fitting procedure. Finally, the Monte Carlo fitting package non-linear least-squares minimization and curve-fitting for Python \citep[LMFIT;][]{LMFIT} that utilises the Markov chain Monte
Carlo (MCMC) Goodman-Weare algorithm python package {\em emcee} \citep{MCMC2013} is used to find the best solution.

We adopt wide parameter spaces of $p=1-5$, $n_c=10^3-10^9$\,cm$^{-3}$ and $r_c=0.0001-0.15$\,pc. Despite of the strong parameter degeneracy, the use of the MCMC approach still allows us to find a best core density model. However, also because of the serious parameter degeneracy, the solution is far from unique and we still have a large room to adjust the density model parameters to improve the fitting of the molecular emission maps by our chemical model in later steps.

\begin{table}
	\centering
	\caption{Fitted density model parameters of the four cloud cores from 2-D column density map fitting ({\em initial core density model}) and the adjusted ones ({\em adjusted core density model}).}
	\label{tab_fits_2d}
	\begin{tabular}{cccccccc}
		\hline
		&&\multicolumn{3}{c}{{\em initial core density model}} & \multicolumn{3}{c}{{\em adjusted core density model}} \\
		&Center coordinate &	$p$    & $n_c$              & $r_c$  &	$p$    & $n_c$              & $r_c$  \\
		&(J2000)    &	       & (cm$^{-3}$)        & (pc)   &	       & (cm$^{-3}$)        & (pc)  \\
		\hline
		C1 & 07:10:05.7, -10:31:10.7 & 3.6  & 3.2$\times 10^8$ & 0.004  & 3.6  & 3.2$\times 10^8$ & 0.004 \\
		C2 & 07:10:06.2, -10:32:00.3 & 4.3  & 5.9$\times 10^6$ & 0.020  & 4.3  & {\bf 4.4$\times 10^6$} & 0.020\\
		C3 & 07:10:05.3, -10:31:36.7 & 1.8  & 2.8$\times 10^6$ & 0.006  & 1.8  & 2.8$\times 10^6$ & 0.006\\
		C4 & 07:10:07.3, -10:31:57.7 & 3.1  & 1.9$\times 10^5$ & 0.024  & {\bf 1.5}  & {\bf 1.0$\times 10^5$ }& {\bf 0.036 }\\
		\hline
	\end{tabular}	\\
	Note: The other two model parameters are fixed to $n_{\rm b}=10^4$\,cm$^{-3}$ for cores C1 and C2, but 0 for C3 and C4. The outer radius is $R=0.17$\,pc for all cores. Adjusted parameters are marked with boldface. To avoid misleading, the fitted uncertainties are not shown because of the strong parameter degeneracy.
\end{table} 

The best model parameters are given in Table~\ref{tab_fits_2d} and the best fit column density map is compared to the observed one in Fig.~\ref{fig_fits_2d}. 
The model map agrees well to the observed one at the positions of cores C1, C2 and C3. There are slightly larger differences in the peripheral regions of C1 and C2 because our core models are all round while the observed ones are slightly elliptic. The differences between the modeled and observed maps are within ratios of 0.75-1.25 (about $\pm 25$\% ), which is small compared to the dynamical range of the observed map (e.g., the column density ratios between the core peaks and the background reach $\sim 30=1.2\times 10^{23}$\,(cm$^{-2}$)$/4\times 10^{21}$\,(cm$^{-2}$)). 

However, the chemically important core C4 is badly fitted because it is strongly affected by the neighboring core C2 in the model fitting. We have done a lot of experiments, together with the chemical simulations that will be mentioned in detail in Sect.~\ref{sec_comp}, to enhance the contribution of C4 to the \ce{H2} column density map and to achieve a good fit to the observed \ce{CCS} and \ce{HC3N} maps at the same time. Eventually, a reasonably good solution for C4 is found to be $r_c=0.036$\,pc $n_c=10^5$\,cm$^{-3}$ and $p=1.5$. This makes C4 the most diffuse core among the four. However, to balance the impact of the modification to the C4 parameters, the neighboring core C2 needs to be refined accordingly. This is achieved by fixing the other three cores and all other parameters of C2 as in Table~\ref{tab_fits_2d} but re-fitting its center density $n_c$ using LMFIT as above. The best fit value turns out to be $n_c=4.4\times 10^6$\,cm$^{-3}$ and the best fit model column density map is compared to the observed one in Fig.~\ref{fig_reasonable_phy}. The middle panel of the figure shows that the new C4 model can better reproduce the extended structures to the east of C2 than in Fig.~\ref{fig_fits_2d}. The  observation-to-model ratio map in the right panel has changed a little bit around C2 but is still within the range of $\pm 25\%$. Hereafter, we will call this {\em adjusted core density model}, while the original model is the {\em initial core density model}. All parameters of these two physical models are listed in Table~\ref{tab_fits_2d}. 

\subsection{Other physical parameters}\label{sec_otherParameters}

In addition to the density structures, we also need radial profiles of gas and dust temperature to model the chemistry in the cloud cores. The gas and dust temperatures are not well constrained by observations for the cores in G224.4-0.6. \citet{Planck2016} reported dust temperatures ranging from 5.8 to 20 K for all the 13188 PGCCs. In this work, we adopt the parametric formula of \citet{Hocuk2017} to express the dust temperature as a function of extinction ($A_{\rm V}$):
\begin{equation}
\label{eq_T}
T_d(A_{\rm V})=\left[11 + 5.7\tanh(0.61-\log(A_{\rm V}))\right]\chi_{\rm uv}^{1.0/5.9}
\end{equation}
where $\chi_{\rm uv}$ is the Draine UV field strength \citep{Draine1978}.  We adopt the typical interstellar UV field with $\chi_{\rm uv}=1.0$. 

The visual extinction can be estimated from the observed \ce{H2} column density along the lines of sight via the empirical formula of \citet{Guver2009}: 
\begin{equation}
A_{\rm V}=N(\rm H_2)/2.2\times 10^{21}\,({\rm mag}).
\end{equation}
As a simple check, Eq(\ref{eq_T}) gives $T_d\sim 6$ and $\sim 13$K for the core center and edge (at a radius of 0.17\,pc), respectively, for the two major cores C1 and C2. Such $T_d$ values are within the temperature range of all PGCCs and typical for cold cores. Considering the beam size of 0.074\,pc, the central dust temperature from our core model within one beam is about $9\sim 10$\,K which is comparable to 11\,K, the value used in Eq.(\ref{eq_NH2}).

The gas temperature is also uncertain in these cores. \citet{Tate2017} deduced kinetic temperature from \ce{NH3}\,1-1 to be $15.5\pm 17.7$\,K at a position $\alpha({\rm J}2000)=$07h10m03.8s, $\delta({\rm J}2000)=$-10d31m28.21s in the cloud that is not at the peak of any of the four cores and the uncertainty is very large. We have to roughly assume the gas temperatures to be equal to that of the dust at all positions, which means that the gas and dust are well thermally coupled everywhere.

We have noted that Eq.(\ref{eq_T}) is a bit inconsistent with the adoption of a constant dust temperature of 11.1\,K in Eq.~\ref{eq_NH2} and the dust and gas temperatures are also not necessarily equal in real cloud cores (particularly in the low density regions). However, the effects of the small temperature differences will not alter our major conclusions (see the discussions in  Sect.~\ref{subsec:rob_uniq}).

\subsection{Chemical model}
\label{sec_chem}
We use a gas-grain chemical model to couple the physical structures to study the molecular chemistry in G224.4-0.6. The model includes gas-phase and dust surface reactions which are linked through the accretion of species and thermal and cosmic-ray induced desorption \citep[see e.g.][]{hase1992,seme2010}. The reaction network was taken from \citet{seme2010} that has included most of the chemical species and reactions relevant to dark clouds. The photo-electron ejection from dust grains and ion accretion were added self-consistently by \citet{Ge2016a} with extended range of grain charges from -5e to +99e. 

We use the FORTRAN chemistry code "GGCHEM" that was successfully benchmarked against the standard models of \citet{seme2010} and was updated to investigate the chemistry effects of turbulent dust motion \citep{Ge2016a} and the chemical differentiation of different dust grain size distributions \citep{Ge2016b}. In this code, we use typical dust grain radius of $0.1\,\mu$m and bulk density of 3.0\,g\,cm$^{-3}$. The ratio of diffusion-to-binding energy of species is fixed as the typical value of 0.5. The cosmic ray ionization rate is set to be $1.3\times 10^{-17}$\,s$^{-1}$. 
The initial abundances are listed in Table~\ref{tab_iab}  which have been widely used as input for chemical models of dense clouds \citep[e.g.][]{Wakelam2008,Hincelin2011,Furuya2011,Majumdar2017}. The dust-to-gas mass ratio is set to the widely used value of 0.01 which had been verified by recent observations \citep[e.g., $\sim 0.011$ in $\rho$\,Oph\,A;][]{Liseau2015}.

The gas-grain chemical model is a single-point model (0-D). We will construct a chemical model for a 1-D core by using a collection of co-evolving 0-D chemical models at different radii.  
We set 10 linearly spaced radial grids for each core, which is more than the number of pixels covered by each core on the observed molecular line maps (the left column of Fig.~\ref{fig_model26}). This 1-D computation populates a 3-D chemical distribution for each core by symmetry. Linear interpolation is used when projecting the 3-D distribution onto the 2-D sky plane. This initial molecular column density maps are cast on a spatial grid on the sky plane that was used for the \ce{H2} column density analysis in Figs.~\ref{fig_fits_2d} and \ref{fig_reasonable_phy}, which is finer by a factor of about 2 than the spatial grids of the observed molecular line maps. All the cores are assumed to be evolving coevally. Finally, all the model molecular-line column-density maps are smoothed by a round Gaussian beam of the same size $18.8''$ as the observations.

\begin{table}
	\centering
	\caption{Initial abundances.}
	\label{tab_iab}
	\begin{tabular}{llllll}
		\hline
		species            &   $n_i/n_{\rm H}$      & species            &   $n_i/n_{\rm H}$ \\
		\hline
		H$_2$              &   0.5                  & S$^+$              &   $8.00\times 10^{-8}$\\
		He                 &   $9.00\times 10^{-2}$ & Fe$^+$             &   $3.00\times 10^{-9}$\\
		C$^+$              &   $1.20\times 10^{-4}$ & Na$^+$             &   $2.00\times 10^{-9}$\\
		N                  &   $7.60\times 10^{-5}$ & Mg$^+$             &   $7.00\times 10^{-9}$\\
		O                  &   $2.56\times 10^{-4}$ & Cl$^+$             &   $1.00\times 10^{-9}$\\
		Si$^+$             &   $8.00\times 10^{-9}$ & P$^+$              &   $2.00\times 10^{-10}$\\
		\hline
	\end{tabular}\\ [1mm]
\end{table}

\section{A successful interpretation to all the CCS, \ce{HC3N} and \ce{N2H+} maps}\label{sec_comp}

Because we mainly care about the positions of the molecular line emission peaks and the geometry of the emission regions, precise radiation transfer of the modeled lines is unnecessary. We thus adopt optically thin conditions for all the three molecular lines so that the molecular column density maps, after being normalized by their strongest pixel, can be directly compared to the observed maps similarly normalized with assumption of constant temperature within the maps. The optically thin assumption can be verified through the analysis of the strongest \ce{N2H+} line, which shows that its optical depth is about 0.9$\pm$0.4 at the SCUBA-2 peak \citep[][]{Tate2017}. The weaker CCS and \ce{HC3N} lines should be optically thin.

We mainly constrain our model age using molecular peak positions and peak column densities. To do this, we calculated the distance ($D$) of molecular peaks between observation and model. We take the half beam size of the molecular line maps (0.05\,pc, the effective spatial resolution) as a reference value to facilitate comparison. The molecular peak column density ($N^{\rm peak}$) will also be considered to constrain the model. To consider both $D$ and $N^{\rm peak}$, we define a relative difference (RD) function as: 
\begin{equation}
{\rm RD}=\sum_{i=1}^{3}\left[\frac{\left|N_{\rm mod,i}^{\rm peak}(t)-N_{\rm obs,i}^{\rm peak}\right|}{N_{\rm obs,i}^{\rm peak}} + \left( \frac{D_i(t)}{\theta_{\rm HPBW}}\right) \right]
\label{eq_RD}
\end{equation}
where $N_{\rm obs,i}^{\rm peak}$ and $N_{\rm mod,i}^{\rm peak}$ are the observed and modeled peak column densities of molecule $i$. The $\theta_{\rm HPBW}=18.8''$ is used to normalize the weighting of the two terms. The smaller the RD value, the better the agreement. Thus the best chemical age will be reached by searching for the minimum RD$_{\rm min}$.

For the observed column densities ($N_{\rm obs}$), we roughly use $N_{\rm obs}^{\rm peak}({\rm CCS})=5\times 10^{12}$\,cm$^{-2}$ \citep[$\sim 10^{12}-10^{13}$ from][]{Tate2017}, $N_{\rm obs}^{\rm peak}({\rm HC_3N})=10^{13}$\,cm$^{-2}$ (assumed), and $N_{\rm obs}({\rm N_2H^+}^{\rm peak})=5\times 10^{13}$\,cm$^{-2}$ \citep[$\sim 10^{13}-10^{14}$ from][]{Tate2017} for the normalization in RD. We have tested that changes of the peak column densities within one order of magnitude do not change the best chemical age corresponding to the RD$_{\rm min}$, especially for \ce{HC3N}.

\subsection{Chemical model with the initial core density model}
\begin{figure*}
	\centering
	\includegraphics[width=0.7\linewidth]{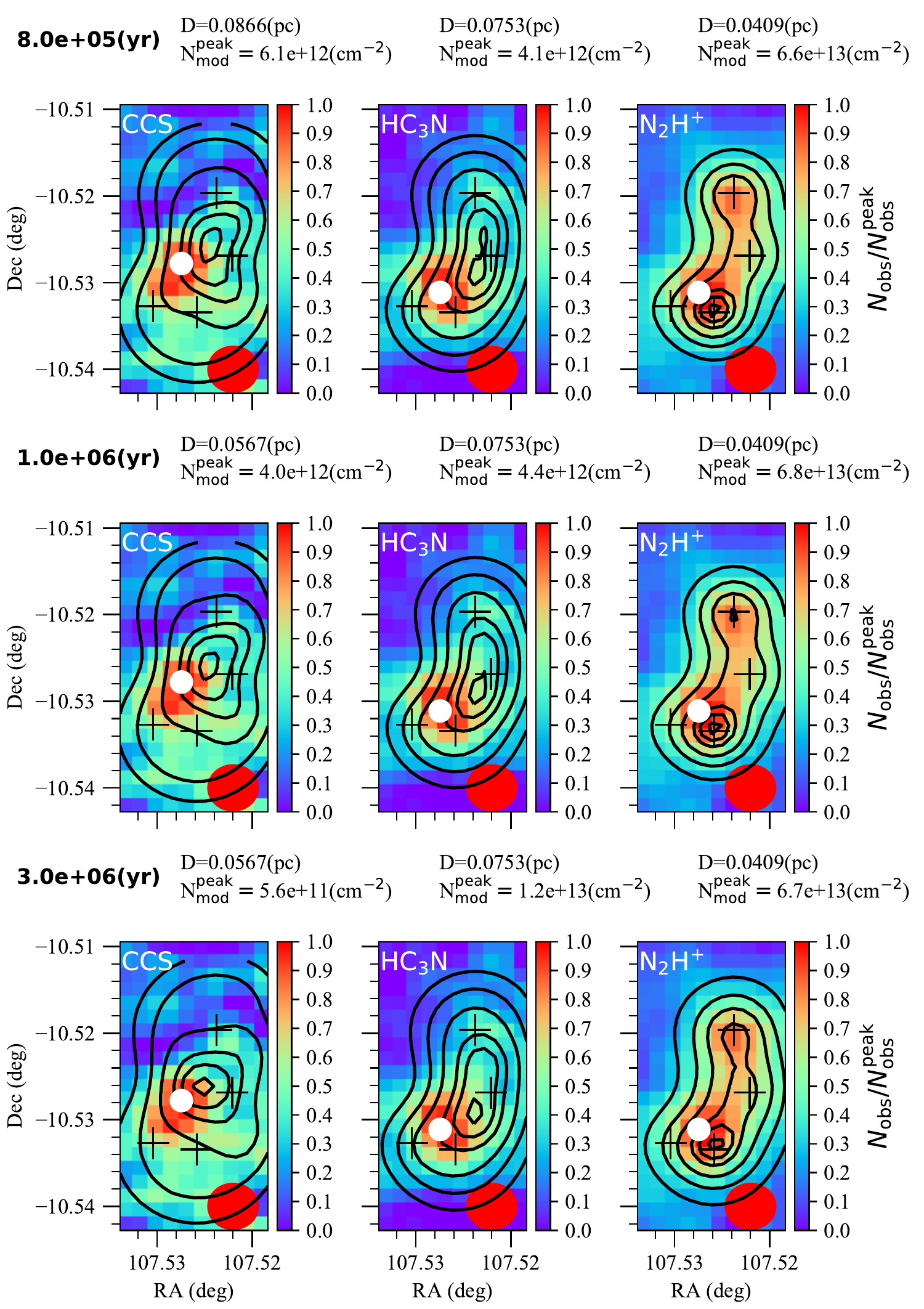}
	\caption{Comparison between normalized (by peak value) observed (color scale) and modeled column density maps (black contours with levels of [0.2, 0.4, 0.6, 0.8, 0.9, 0.98]$\times N_{\rm peak}$) of CCS (left panels), \ce{HC3N} (middle panels) and \ce{N2H+} (right panel) from {\em initial density core model}. The $N_{\rm mod}^{\rm peak}$ value, model age and distance ($D$) between observed and modeled molecular peaks are shown with labels over each panel. The black crosses show the four-core center positions. The white point marks the observed molecular peak position. The red circle indicates beam size. See details of observed peak column densities in text.}
	\label{fig:comparison26}
\end{figure*}

\begin{figure}
	\centering
	\includegraphics[width=0.7\linewidth]{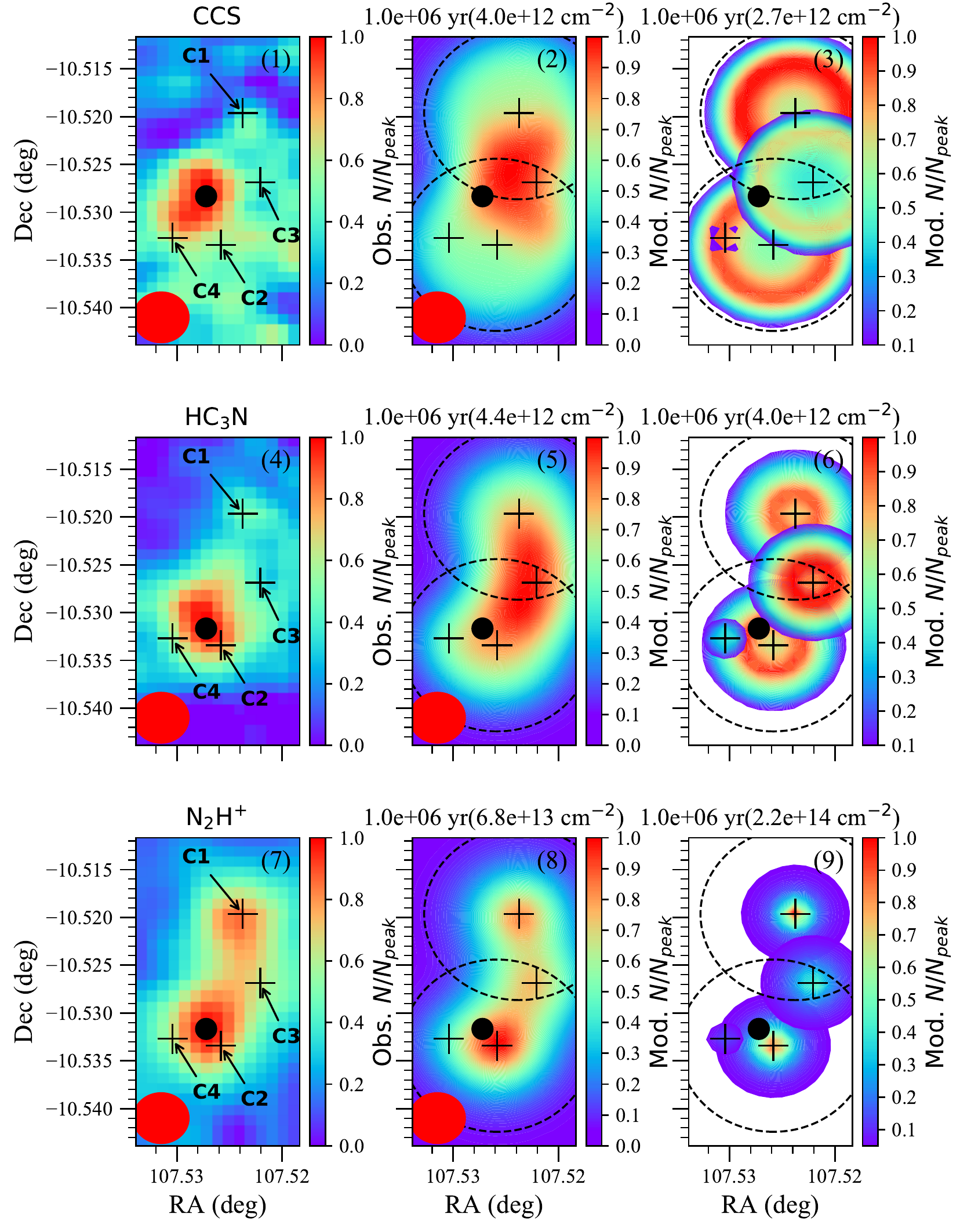}
	\caption{Comparison between observed (left column) and modeled (middle column) maps of CCS (top row), \ce{HC3N} (middle row) and \ce{N2H+} (bottom row) at age of $10^6$\,yr, using the {\em initial core density models} in Table~\ref{tab_fits_2d}. The right column shows the individual contribution of each core to the total column density maps in the middle column (the order of overlapping does not matter because of the optically thin conditions). The pixel values (shown with color scale) of each plot are always normalized by the strongest pixel in that plot whose absolute value is given in the title of each panel. A lower cutoff of the color scale (the lower limit of the color bar) in panels (3), (6) and (9) is used to better visualize the contribution of individual cores. Black crosses indicate the center positions of the four cores. The observed molecular peak position is marked by a black point. The beam sizes are indicated in the bottom-left corners for both observed and modeled maps. The large dashed-line circles in the middle and right columns show the outer radius $R$ of the two major cores C1 and C2. }
	\label{fig_model26}
\end{figure}

\begin{figure*}
	\centering
	\includegraphics[width=1.0\linewidth]{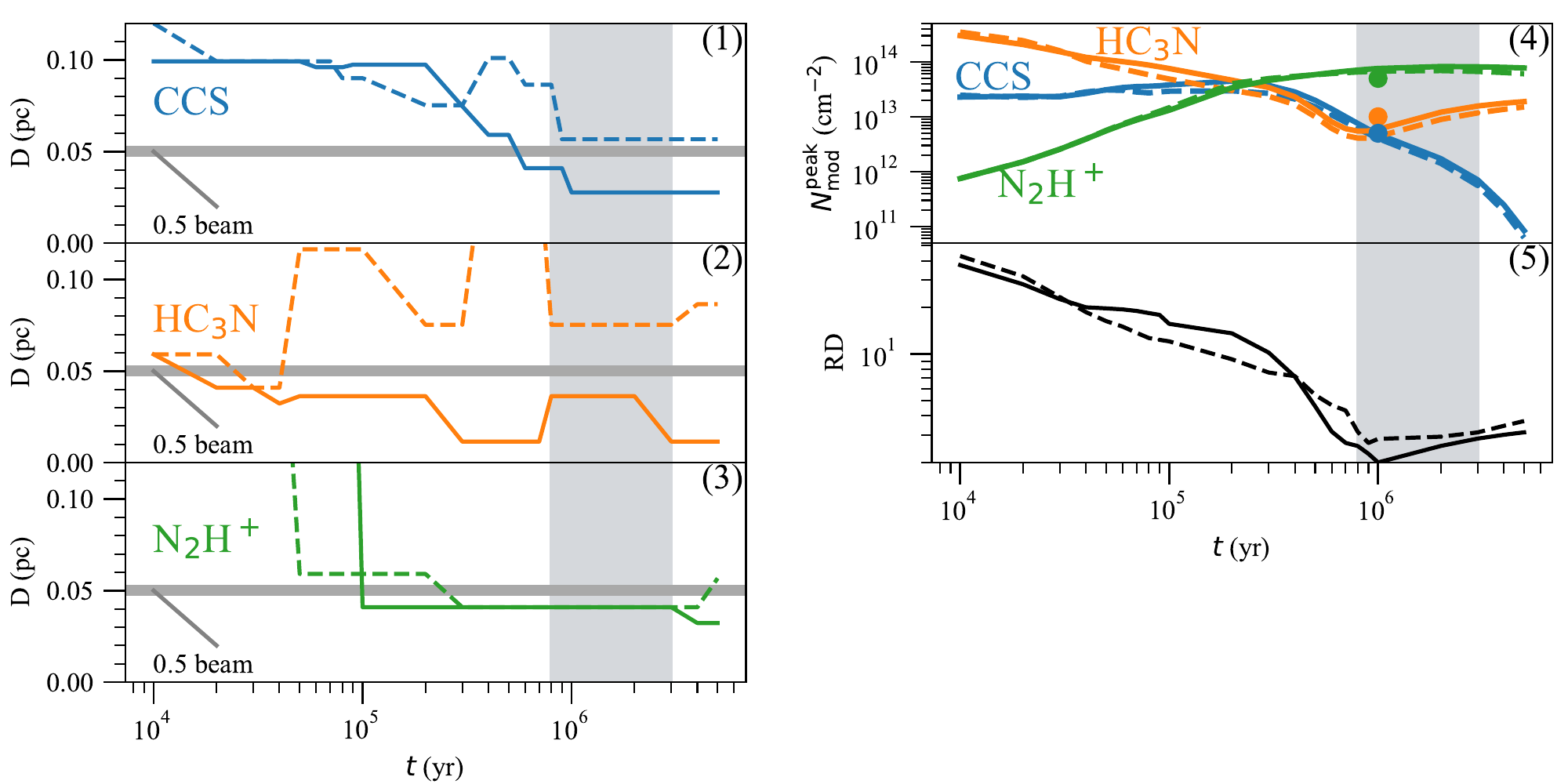}
	\caption{Distances ($D$) between modeled and observed molecular column density peaks of CCS (blue, panel (1)), \ce{HC3N} (orange, panel (2)) and \ce{N2H+} (green, panel (3)) from the {\em initial core density model} (dashed lines) and {\em adjusted core density model} (solid lines). In panel (1-3), the gray line indicates the half of beam size of 0.05\,pc to facilitate comparison. Panel (4): peak column densities from {\em initial core density model} (dashed lines) and {\em adjusted core density model} (solid lines). The points indicate the observed column densities we used for comparison which are placed at $10^6$\,yr. Panel (5): Relative difference (RD) between observation and model constrained by Eq(\ref{eq_RD}). In all panels, possible age range of $8\times 10^5$ to $3\times 10^6$\,yrs is indicated by the vertical light gray region.}
	\label{fig:DN}
\end{figure*}

We first run the chemical code for the {\em initial core density model} in Table~\ref{tab_fits_2d}. The molecular column density maps from observations (color scale) and models (contour) at selected ages of $8\times 10^5$, $10^6$ and $3\times 10^6$\,yrs are shown in Fig.~\ref{fig:comparison26} together with calculated distance ($D$) of molecular peaks between observed and modeled ones shown with labels. 
The peak distances ($D$), column densities ($N_{\rm mod}^{\rm peak}$) and relative difference (RD) as function of model age are shown in Fig.~\ref{fig:DN} for comparison. By checking the $D$ values (see dashed lines in left panels of Fig.~\ref{fig:DN}), we found that the differences are typically larger than the reference value of $D=0.05$\,pc for CCS and \ce{HC3N} before $8\times 10^5$\,yr. Especially for \ce{HC3N}, the $D$ value is up to $\sim 0.08$\,pc, about 0.8 beam. After $3\times 10^6$\,yr, the CCS column density drops sharply to be smaller than $10^{12}$\,cm$^{-2}$. Thus the best chemical age is constrained to be $\sim 10^6$\,yr by RD$_{\rm min}$ (dashed line in panel (5) of Fig.~\ref{fig:DN}) at which the molecular column densities are comparable to the observed ones \citep{Tate2017}. The minimum of the RD curve in the figure is so shallow that the solution is similarly good in a range of chemical age of $8\times 10^5\sim 3\times 10^6$. Even within this age range, the $D$ values of CCS and \ce{HC3N} are still larger than 0.05\,pc, which hints that the {\em initial core density model} fails to reproduce observed molecular peaks even at the possible best chemical age of $10^6$\,yr. 

From Fig.~\ref{fig:comparison26} at the selected repesentative ages of $8\times 10^5$ (top row), $10^6$ (middle row) and $3\times 10^6$\,yrs (bottom row), we can see that the modeled CCS peak is still on the symmetry line of the two major cores C1 and C2, while \ce{HC3N} shows a ridge that offsets toward the core C3, with both being opposite to the observed offset of its peak.
To help understand the origin of the discrepancies, the column density maps of CCS, \ce{HC3N} and \ce{N2H+} at the best age ($10^6$\,yr) are shown together with the contribution of each individual cores (the contribution plot; right panel) in Fig.~\ref{fig_model26}. 
The contribution plots in panels (3) and (6) tell us that the CCS peak is mainly contributed by the CCS shells of the major cores C1 and C2, while the \ce{HC3N} ridge is caused by the core C3.

The modeled peaks of both CCS and \ce{HC3N} need to be shifted toward the east direction to agree with the observations. The most straightforward way we can learn from the geometrical relationship of the chemical distributions in this model is to enhance the contribution of the core C4 which is currently too small. Therefore after some rounds of iterations between the manual adjustment of core density parameters (for C2 and C4) and the comparison of the resulting chemical models with the  observations, a good set of core-density-model parameters of C4 (and also C2) have been found as described in Sect.~\ref{sec_fits_2d}; this constitutes the {\em adjusted core density model} which we will explore in the next subsection.  

\subsection{Chemical model with the adjusted core density model}
\label{sec_reasoanble_case}
\begin{figure*}
	\centering
	\includegraphics[width=0.65\linewidth]{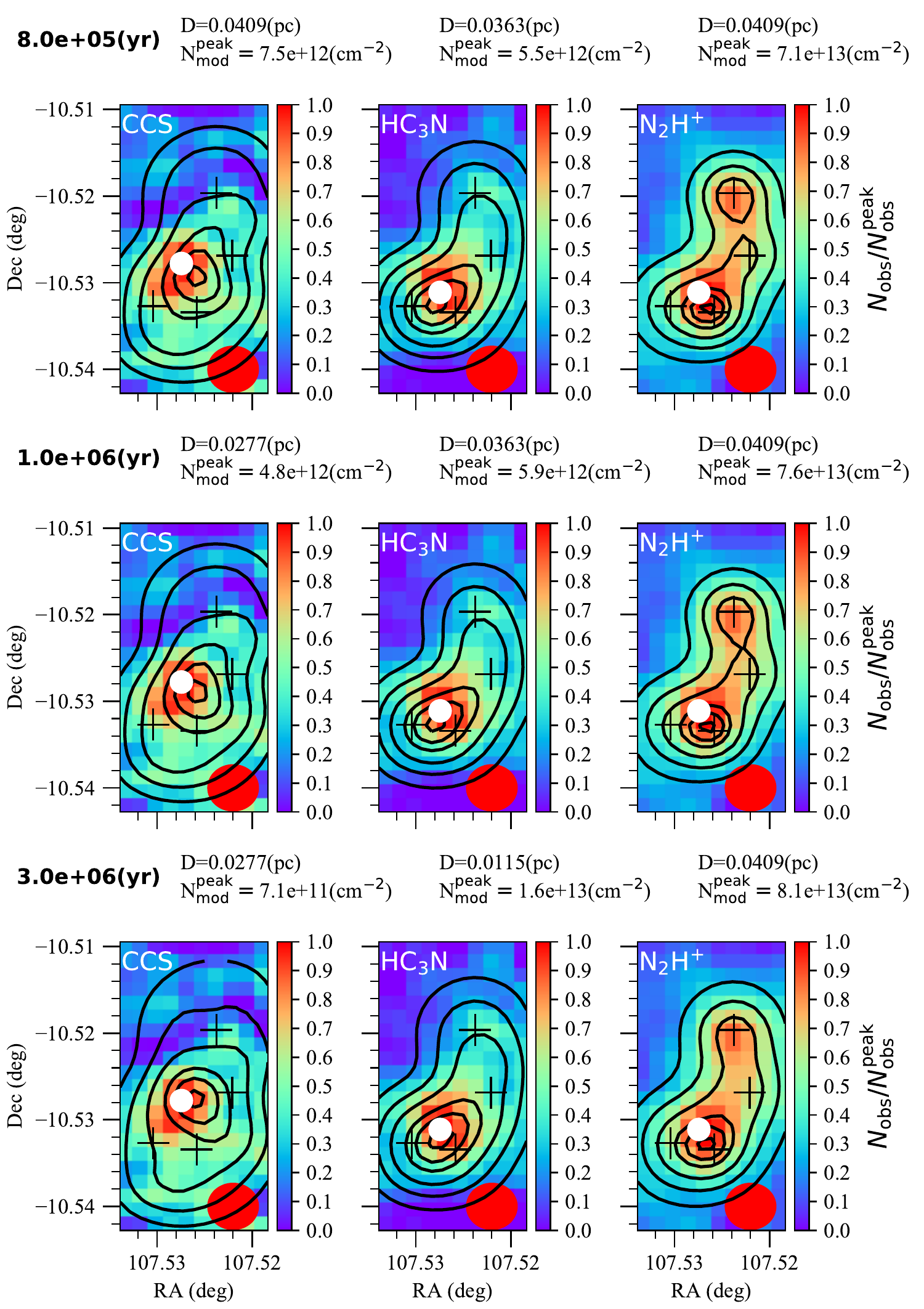}
	\caption{Same as Fig.~\ref{fig:comparison26}, but with {\em adjusted core density model}.}
	\label{fig_reasonable}
\end{figure*}

With the {\em adjusted core density model}, we find a significantly better agreement between our chemical model maps and the observed maps of all three molecules CCS, \ce{HC3N} and \ce{N2H+} at the best age range of $8\times 10^5\sim 3\times 10^6$\,yrs, as shown in Fig.~\ref{fig_reasonable}. The chemical age range is jointly constrained by the peak positions (the distances $D$) of CCS and \ce{HC3N} which are now both coincident within a half beam (0.05 pc) (see solid lines in left panels of Fig.~\ref{fig:DN}). Meanwhile the RD$_{\rm min}$ is also reached at $10^6$\,yr. The modeled peak column density of \ce{N2H+} ($\sim 6.8\times 10^{13}$\,cm$^{-2}$) is also in the range of the observed values of $\sim 7.7\times 10^{12}\sim 1.4\times 10^{14}$\,cm$^{-2}$ from \citet{Tate2017}. An optical depth of $\tau=1.0$ may result in an underestimation of the observed column density by a factor of $\sim 1.6$ ($=1/(1-\exp(-1))$). The optical depth corrected modeled column density of \ce{N2H+} is $\sim 1.1\times 10^{14}$\,cm$^{-2}$ which still agrees to the observed value.

\begin{figure*}	
	\includegraphics[width=1.0\linewidth]{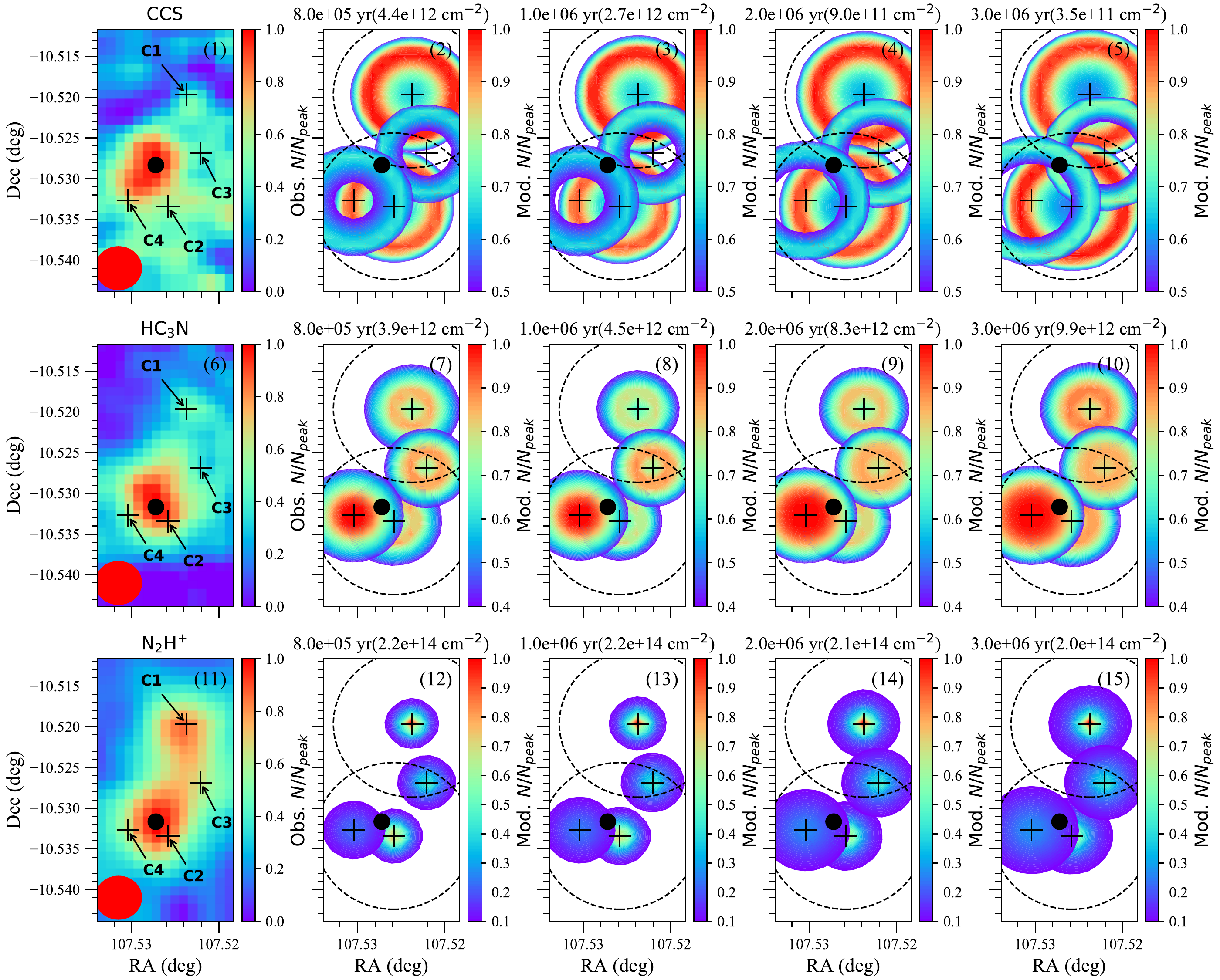}
	\caption{Same as the left and right columns of Fig.~\ref{fig_model26} but for the case using the {\em adjusted core density model} at different chemical ages shown in panel titles.}
	\label{fig_reasonable_deconv}
\end{figure*}

The contribution plots in Fig.~\ref{fig_reasonable_deconv} help us understand how the above good agreements are achieved. Because of the depletion effect, the CCS column density map of each individual core is a ring band. The key difference from the results of the {\em initial core density model} in Fig.~\ref{fig_model26} is the enhanced CCS ring of the core C4. The C4 ring first pull the CCS emission peak in Fig.~\ref{fig_reasonable} to the east side through its overlap with the rings of C2 and C3 around the age of $\sim 8\times 10^5$\,yr (panel 2 of Fig.~\ref{fig_reasonable_deconv}). Then, the sizes of the CCS rings of all four cores increase with time (panels 3-5 of Fig.~\ref{fig_reasonable_deconv}) and the CCS emission peak moves upward  (left panels of Fig.~\ref{fig_reasonable}) due to the change of the overlap effects. Similarly, the \ce{HC3N} column densities also consist of ring bands around all four cores, except that the ring sizes are smaller. The offset of the \ce{HC3N} emission peak towards the north-east direction in the middle column of Fig.~\ref{fig_reasonable} is mainly supported by the overlap of the \ce{HC3N} rings of the cores C2, C3 and C4 in the middle row of Fig.~\ref{fig_reasonable_deconv}. Interestingly, the \ce{N2H+} emission peak in our model (right panels of Fig.~\ref{fig_reasonable}) also has shifted slightly toward the north-east direction as observed. This is again the consequence of core overlapping among C2, C3 and C4 (panels 12-15 of Fig.~\ref{fig_reasonable_deconv}), though they do not develop any depletion ring.

\section{Discussion}\label{sec_discussions}
\subsection{Involved chemistry}\label{sec_chemistry}

\begin{figure*}
	\centering
	\includegraphics[width=1.0\linewidth]{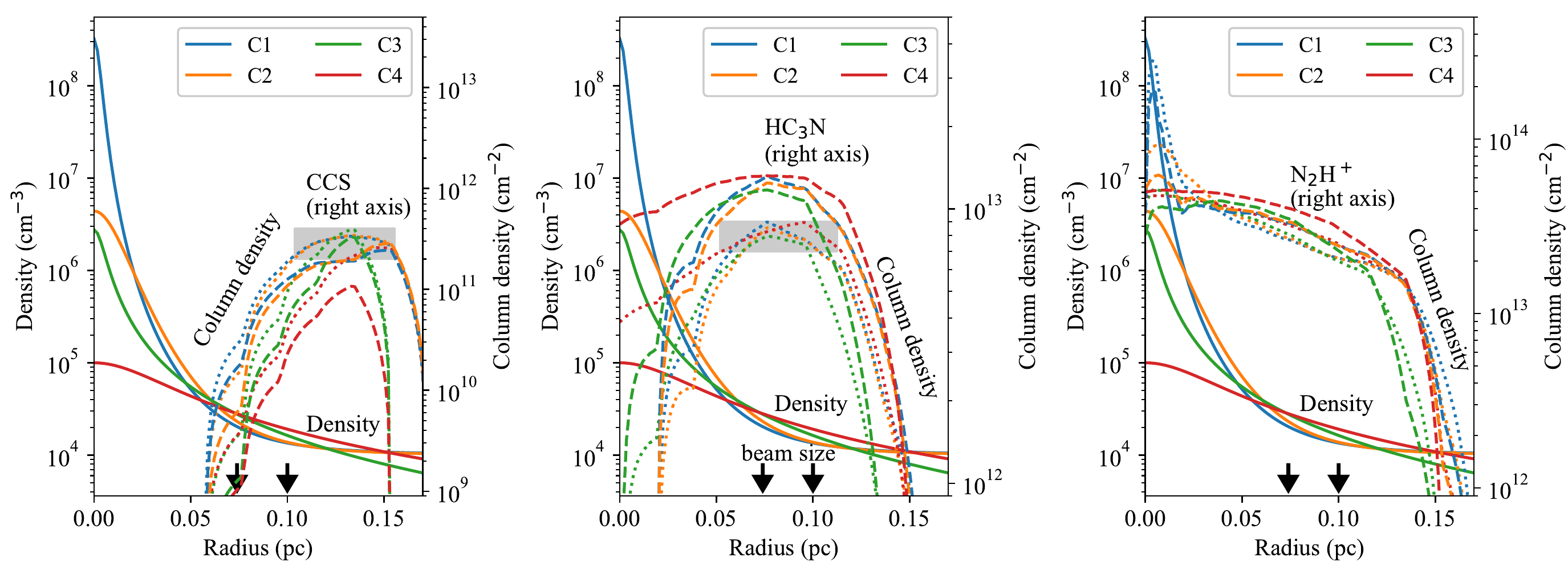}
	\caption{Radial profiles of gas density (solid lines; left axes) and molecular column  density (dotted lines, right axes) in our best chemical model (at the age of $10^6$\,yr) constructed on the basis of the {\em adjusted core density model} (see in text). Dashed lines are similar as the dotted lines but for a model with the gas and dust temperatures forced to 15\,K everywhere. The gray shaded regions mark the rough positions of the shells resulted from core-center depletion, which correspond to radii of $r=0.104-0.156$\,pc for CCS (left panel) and $r=0.052-0.112$\,pc for \ce{HC3N} (right panel). The arrows indicate the beam sizes of the continuum (0.074\,pc) and molecular column density (0.1\,pc) maps.}
	\label{fig_profiles}
\end{figure*}

The overlapping effects of molecular shell/ring patterns of CCS and \ce{HC3N} is mainly driven by their depletion at the core center due to adsorption onto dust grains, with some help from the photo-dissociation effects at the boundary. We compare the radial profiles of the CCS and \ce{HC3N} column densities with the density profiles of all four cloud-core models at the best age of $10^6$\,yrs in Fig.~\ref{fig_profiles}. Although density profiles are very different in the center of the four cloud cores, they are just similar at the outer radii of the cores that are resolved by the telescope beam. This could be a natural consequence of the need to fit the chemical patterns. That is to say, only when the densities in the peripheral regions of the four cores are similar, the chemical models are able to generate ring patterns with comparable column densities for the considered molecules to reproduce the observed line emission patterns through the overlapping effects.

The most salient difference between CCS and \ce{HC3N} is the different radial positions of their column density peaks. It is clear in both Fig.~\ref{fig_profiles} and previous model maps that the CCS rings are larger than that of \ce{HC3N} in all four cores at the best chemical age of $10^6$\,yrs. This is due to the shorter depletion time scale of the former than that of the latter. These molecules have similar time scales of about $10^6$\,yrs for their adsorption onto grains at the gas density of about $10^4$\,cm$^{-3}$ and gas and dust temperatures of about 10\,K and dust-to-gas mass ratio of 0.01 \citep[e.g.][]{Leger1985} at the positions of the ring structures. However, \ce{HC3N} is closely linked to the volatile molecule \ce{N2} \citep[e.g.][]{Bergin1997,Charnley1997}, which makes the time scale of its depletion from the gas phase significantly longer than that of CCS because the gas-phase \ce{N2} has a high abundance of $\sim 10^{-5}$ which delays the depletion of the gas phase \ce{HC3N} via  \ce{N2}$\rightarrow$CN$\xrightarrow[]{\ce{C2H2}}$\ce{HC3N}.

\subsection{Robustness and uniqueness of the model}\label{subsec:rob_uniq}
Because of the limitation of available data of our target G224.4-0.6, we only briefly test the robustness of our four-cores chemical model against the velocity information in the observed line maps and the dust and gas temperatures derived from \ce{NH3} line and Herschel data. The observed moment-1 maps (not shown in this work) of \ce{N2H+} \citep[see also the Fig.17 of ][]{Tate2017}, \ce{HC3N} and CCS indicate that the differences in average line-of-sight velocities are no larger than about 0.5, 0.5 and 0.9\,km\,s$^{-1}$ for the three molecules respectively among the positions of the four model cores and the three molecular peaks, which is not too large compared to the typical line width of $\sim 1.46$\,km\,s$^{-1}$ for \ce{N2H+} and \ce{HC3N}, and a larger width of $\sim 2-3$\,km\,s$^{-1}$ for CCS.
Therefore the overlapping effects of the four cores can work as expected in the chemical model.

The dark cloud chemistry is sensitive to gas and dust temperatures. As we have mentioned in Sect.\ref{sec_otherParameters}, the gas and dust temperatures of our target are uncertain and an empirical radial temperature profile is adopted in the modeling. \citet{Tate2017} constrained a very rough gas temperature of $15.5\pm 17.7$\,K from the \ce{NH3}\,(2,2) line observed with a large beam of $74''$ towards a position in the peripheral of the cores. In addition, the WISE 22\,$\mu$m image of G224.4-0.6 \citep[see Fig.13 of][]{Tate2017} has shown IR emission at the positions of our model cores 1, 2 and 3, indicating possible heating of the cloud cores from inside by embedded protostars. The Herschel dust temperature map of this region \citep{Marsh2017} also shows higher dust temperatures in the range of $\sim 12-16$\,K. To test the robustness of our four-cores model against the variation of temperature, we have re-computed our best model (the {\em adjusted core density model}) with a constant radial dust and gas temperatures of 15\,K for all the four cores (this is reasonable to model the low resolution maps in this work). The modeled radial molecular column density profiles are shown with dashed lines in Fig.~\ref{fig_profiles}, from which we see that this change of temperature indeed affects the abundances, but within a small factor of $\sim 2$. Our chemical ring overlapping mechanism still works, because the radii of the rings are similar as in our original best model (the shaded region in the figure).

Although our four-cores chemical model has reproduced all available observations well, the lack of uniqueness of the model is an important weakness to be improved in near future. Beside the non-unique solutions of the core density models, as mentioned in earlier subsections, there are more aspects to improve from both observational and modeling angles. For example, the SCUBA-2 map could have missed a smooth structural component of the cloud due to spatial filtering, which could affect the peripheral region of our core models by adding an additional extinction and/or contributing additional abundances from the diffuse gas that could be still missing in our models. Mutual shielding of interstellar UV light among the four cores may also have some effects to the modeled chemical patterns, if the four cores are close enough to each other; Non-uniform illumination of interstellar radiation field, as suggested for the starless core L1544 by \citet{Spezzano2016} may also has some impact to the chemistry in the peripheral regions of the cores. For another example, our chemical model also adopts a static physical model for each core and all four cores start the chemical evolution from the same time spot, while the real cores might started their formation from different moments and collapse at different dynamical timescales. It is still an open question whether alternative solutions are possible when more such model components are taken into account and when more observational evidences become available.

\subsection{Spatial projection effects on interstellar chemistry}\label{sec_coreOverlap}

The overlapping effect of two or more cloud cores explored in this work is to offset the molecular column density peaks away from the \ce{H2} column density peaks. Such offsets of molecular line peaks are ubiquitous in observed cloud maps \citep[see, e.g.,][]{Tate2017,Spezz2017,Seo2019,Nagy2019}. The multiple-cores model provides a viable way to simulate the chemistry of molecules whose lines are not very optically thick in the extended low density regions of the clouds. More opaque lines also can be modeled by adding a radiative transfer treatment. Particularly, molecules that have an extended spatial distributions in the cloud cores, such as the shell structures due to the depletion of CCS and \ce{HC3N} in this work, are easiest to produce observable core overlapping effects. Even for the dense gas tracers that concentrate on the core center, such as \ce{N2H+} in this work, the smaller emission-peak offsets still can be observed by a single dish or interferometer. 

The spatial projection effect on chemistry in a 3D cloud structure is worth of further explorations. More generally, the impact of such 3D models is to break up the confinement to chemical models imposed by the column density map and/or temperature maps and allow the molecular column densities to be built up through the non-local overlapping effect. It offers the freedom to combine structural components that have different sizes, shapes and masses, density and temperature profiles, ambient or internal radiation fields, and even different dynamical evolution histories. Although fully 3D modeling may not be easy or necessary, some simplified versions such as the multiple-core model in this work are still able to capture the most important chemical features in the observed line maps.

\section{Summary}
\label{sec_con}

Offsets of molecular line emission peaks from \ce{H2} column density peaks are very common in observed molecular clouds, which are usually difficult to interpret with a single round core chemical model. We take the example of the Planck cold clump G224.4-0.6 in this work to demonstrate a new approach to interpreting this phenomenon: spatial projection effect of a 3D cloud. 

The gas density structures of G224.4-0.6 is simplified into four individual spherical cores. The core structures are empirically constructed from the SCUBA-2 850\,$\mu$m dust continuum map, named as {\em initial core density model}. Then it is manually fine tuned to achieve much better fittings to both the continuum and molecular emission maps of CCS, \ce{HC3N} and \ce{N2H+}, particularly the asymmetry in the line maps (our adjusted core density model). 

The spatially extended abundance distributions of these molecules in our gas-grain chemical models, particularly the projected ring like patterns due to the depletion of CCS and \ce{HC3N}, overlap with each other to produce enhanced line emission peaks that are generally off the continuum peaks. The best chemical model of our target clump is found within an age range of $8\times 10^5 \sim 3\times 10^6$\,yr. The spatial distributions of the \ce{H2} column density and the spatial offsets of the line emission peaks of all the three molecules are satisfactorily reproduced. This demonstrates the great potential of the spatial projection effect on interpreting the line maps of many other molecular clouds.

\acknowledgments
J.X.Ge acknowledges support from FONDECYT grant 3170768 and the "Light of West China" program of Chinese Academy of Sciences (CAS). 
D.M. acknowledges support from CONICYT project Basal AFB-170002. 
J.H.He thanks the National Natural Science Foundation of China under Grant Nos. 11873086 and U1631237 and the support by Yunnan Province of China (No.2017HC018).
JEL is supported by the Basic Science Research Program through the National Research Foundation of Korea (grant No. NRF-2018R1A2B6003423) and the Korea Astronomy and Space Science Institute under the R\&D program supervised by the Ministry of Science, ICT and Future Planning. 
NI gratefully acknowledges support of CONICYT International Networks for young researchers Grant REDI170243 and CONICYT/PCI/REDI170243 and /REDES190113.
SF acknowledges the support of the EACOA fellowship from the East Asia Core Observatories Association, which consists of the National Astronomical Observatory of China, the National Astronomical Observatory of Japan, the Academia Sinica Institute of Astronomy and Astrophysics, and the Korea Astronomy and Space Science Institute. 
This work is sponsored (in part) by the Chinese Academy of Sciences (CAS), 
through a grant to the CAS South America Center for Astronomy (CASSACA) in Santiago, Chile.

\appendix

\section{Fitting the 1-D column density profiles of the main cores G224NE and G224S}\label{sec_fits_1d}

\begin{figure*}
	\centering
	\includegraphics[scale=0.7]{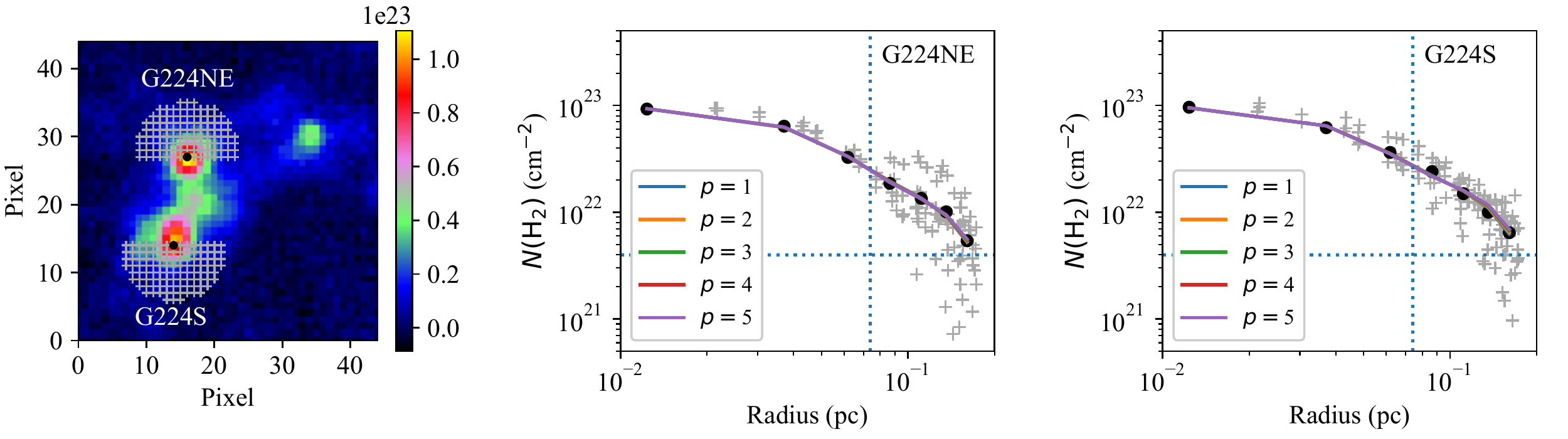}
	\caption{Left panel: Observed \ce{H2} column density map overlaid by pixels within radius of $0.17$\,pc which are used to obtain the bin-averaged column density profiles. Middle and right panels are for the comparison between the extracted bin-averaged \ce{H2} column density radial profiles (black dot) and fitted profiles with fixed $p(=1,2,3,4,5)$ (lines), for G224NE and G224S respectively. The gray crosses show the pixel values. Vertical and horizontal lines indicate the beam size and the background respectively. The solid lines of different $p$ values overlap with each other so closely that they are hardly distinguishable in the middle and right panels.}
	\label{fig_fits_C1C2}
\end{figure*}

We start our testing from fitting 1-D density models to the two major cores G224NE and G224S. The way to construct column density profiles of each core is the same as in the the 2-D fitting case in Sect.~\ref{sec_fits_2d}. To avoid the interference of the profound extended structures between the two cores, we do azimuthal averaging to construct the observed 1-D column density profiles for the two cores only using the image pixels in the half space on the side away from the inter-core direction (i.e., those pixels that make an azimuth angle of $\pi/2$ to $3\pi/2$ with the inter-core direction), as indicated by the gray plus symbols in the left panel of Fig.~\ref{fig_fits_C1C2}. The pixel column densities are shown as a function of radius to the core centers with gray symbols in the middle (for G224NE) and right (for G224S) panels of the figure respectively. Then, each so sampled radial range is evenly divided into seven linedeviationar bins and the average column density in each bin, shown as the black dots in the figure, is used for model fitting. The outer radii are fixed as $R=0.17$\,pc where the \ce{H2} column density roughly drops to the background value of $\sim 4\times 10^{21}$\,cm$^{-2}$ (see Fig.~\ref{fig_fits_C1C2}) estimated from emission-free regions.

Accordingly, we follow the same procedure as in Sect.~\ref{sec_fits_2d} to construct the column density map of each core individually from a given set of 1-D density model parameters. After convolving them with the Gaussian SCUBA-2 beam of 14$''$Spezzano et al. (2016), the 1-D model column density profiles ($N_{\rm H_2}(r)$) are extracted and compared with the observed ones to define the posterior likelihood function: 
\begin{equation}
\prod_i P(y_i|r_i,n_c,r_c,p,n_b,\sigma^2)=\prod_i  \frac{1}{\sqrt{2\pi}\sigma}\exp{\frac{(y_i-N_{\rm H_2}(r_i))^2}{2\sigma^2}},
\end{equation}
where $\sigma$ is the standard deviation of the Gaussian distribution. Finally, the same Monte Carlo fitting method is used to find the best solution.
We similarly adopt wide parameter spaces of $n_c=10^3-10^9$\,cm$^{-3}$, $r_c=0.0001-0.15$\,pc and $n_{\rm b}=0-10^5$\,cm$^{-3}$. However, because of the parameter degeneracy problem with the low spatial resolution map, as already noted by other authors such as \citet{Smith2014} and \citet{Liuhongli2018}, we have to confine ourselves to fixed discrete values of 1, 2, 3, 4 and 5 for the parameter $p$ to ease the search of the best solutions.  
The chosen values of $p$ are representative in other observation works \citep[e.g.,][]{Caselli2002,Tafalla2002,Tang2018}.
\begin{table}
	\centering
	\caption{Fitted density model parameters for G224NE and G224S with various fixed $p$ values.}
	\label{tab_fits_C1C2}
	\begin{tabular}{llll}
		\hline
		$p$    & $n_c$              & $r_c$        &   $n_{\rm b}$ \\
		& (cm$^{-3}$)        & (pc)         &  (cm$^{-3}$)\\
		\hline
		\multicolumn{4}{c}{G224NE} \\
		1     &    4.4e+08 &      4.2e-6 &    2.8e+03\\
		2     &    3.8e+08 &      4.5e-4 &    1.2e+04\\
		3     &    2.2e+08 &      2.6e-3 &    1.5e+04\\
		4     &    5.0e+07 &      7.8e-3 &    1.6e+04\\
		5     &    2.2e+07 &      1.3e-2 &    1.7e+04\\
		\hline
		\multicolumn{4}{c}{G224S} \\
		1 &    4.3e+08 &    4.2e-6 &        6.2e+03\\
		2 &    3.9e+08 &    4.2e-4 &        1.6e+04\\
		3 &    2.9e+08 &    2.1e-3 &        1.9e+04\\
		4 &    1.3e+08 &    5.2e-3 &        2.0e+04\\
		5 &    5.9e+07 &    9.1e-3 &        2.1e+04\\
		\hline
	\end{tabular}	\\
\end{table}

The best fit model parameters are listed in Table~\ref{tab_fits_C1C2}, while the best fit column density profiles (lines with different colors) are over plotted with the observed ones in the middle (for G224NE) and right panels (for G224S) of Fig.~\ref{fig_fits_C1C2}. Although the parameters of the best models vary widely, with the central densities ranging from $\sim 4\times  10^8$ to $\sim 2\times 10^7$\,cm$^{-3}$ and the size of the inner flat core from $\sim 4\times 10^{-6}$ to $\sim 1\times 10^{-2}$\,pc, the best models with different $p$ values agree to each other so closely in the plots that they are almost indistinguishable by eyes. This demonstrates that the SCUBA-2 map resolution (indicated by the vertical dotted lines) is insufficient to differentiate the density models with different $p$ values. However, the best fit background density is similar ($\sim 10^4$\,cm$^{-3}$) in all models except the one with the smallest $p$. The model column densities at the outer edge of each core are all consistently close to and above the observed background column density of $\sim 4\times 10^{21}$\,cm$^{-2}$ (represented by the horizontal dotted line in the plots). Thus, this background density value ($10^4$\,cm$^{-3}$) will be used and fixed for the two major cores in our 2-D fitting in Sect.~\ref{sec_fits_2d} to help reduce the size of parameter space.




%

\bibliography{pgcc} 

\begin{thebibliography}{}
\expandafter\ifx\csname natexlab\endcsname\relax\def\natexlab#1{#1}\fi
\providecommand{\url}[1]{\href{#1}{#1}}

\bibitem[{{Aikawa} {et~al.}(2001){Aikawa}, {Ohashi}, {Inutsuka}, {Herbst}, \&
  {Takakuwa}}]{Aikawa2001}
{Aikawa}, Y., {Ohashi}, N., {Inutsuka}, S.-i., {Herbst}, E., \& {Takakuwa}, S.
  2001, \apj, 552, 639

\bibitem[{{Benson} {et~al.}(1998){Benson}, {Caselli}, \& {Myers}}]{Benson1998}
{Benson}, P.~J., {Caselli}, P., \& {Myers}, P.~C. 1998, \apj, 506, 743

\bibitem[{{Bergin} {et~al.}(2002){Bergin}, {Alves}, {Huard}, \&
  {Lada}}]{Bergin2002}
{Bergin}, E.~A., {Alves}, J., {Huard}, T., \& {Lada}, C.~J. 2002, \apjl, 570,
  L101

\bibitem[{{Bergin} {et~al.}(2001){Bergin}, {Ciardi}, {Lada}, {Alves}, \&
  {Lada}}]{Bergin2001}
{Bergin}, E.~A., {Ciardi}, D.~R., {Lada}, C.~J., {Alves}, J., \& {Lada}, E.~A.
  2001, \apj, 557, 209

\bibitem[{{Bergin} \& {Langer}(1997)}]{Bergin1997}
{Bergin}, E.~A., \& {Langer}, W.~D. 1997, \apj, 486, 316

\bibitem[{{Caselli} {et~al.}(2002){Caselli}, {Benson}, {Myers}, \&
  {Tafalla}}]{Caselli2002}
{Caselli}, P., {Benson}, P.~J., {Myers}, P.~C., \& {Tafalla}, M. 2002, \apj,
  572, 238

\bibitem[{{Caselli} {et~al.}(1999){Caselli}, {Walmsley}, {Tafalla}, {Dore}, \&
  {Myers}}]{Caselli1999}
{Caselli}, P., {Walmsley}, C.~M., {Tafalla}, M., {Dore}, L., \& {Myers}, P.~C.
  1999, \apjl, 523, L165

\bibitem[{{Charnley}(1997)}]{Charnley1997}
{Charnley}, S.~B. 1997, \mnras, 291, 455

\bibitem[{{Claria}(1974)}]{Claria1974}
{Claria}, J.~J. 1974, \aj, 79, 1022

\bibitem[{{Draine}(1978)}]{Draine1978}
{Draine}, B.~T. 1978, \apjs, 36, 595

\bibitem[{{Fiege} \& {Pudritz}(2000)}]{Fiege2000}
{Fiege}, J.~D., \& {Pudritz}, R.~E. 2000, \mnras, 311, 85

\bibitem[{{Foreman-Mackey} {et~al.}(2013){Foreman-Mackey}, {Hogg}, {Lang}, \&
  {Goodman}}]{MCMC2013}
{Foreman-Mackey}, D., {Hogg}, D.~W., {Lang}, D., \& {Goodman}, J. 2013, \pasp,
  125, 306

\bibitem[{{Furuya} {et~al.}(2011){Furuya}, {Aikawa}, {Sakai}, \&
  {Yamamoto}}]{Furuya2011}
{Furuya}, K., {Aikawa}, Y., {Sakai}, N., \& {Yamamoto}, S. 2011, \apj, 731, 38

\bibitem[{{Ge} {et~al.}(2016{\natexlab{a}}){Ge}, {He}, \& {Li}}]{Ge2016b}
{Ge}, J.~X., {He}, J.~H., \& {Li}, A. 2016{\natexlab{a}}, \mnras, 460, L50

\bibitem[{{Ge} {et~al.}(2016{\natexlab{b}}){Ge}, {He}, \& {Yan}}]{Ge2016a}
{Ge}, J.~X., {He}, J.~H., \& {Yan}, H.~R. 2016{\natexlab{b}}, \mnras, 455, 3570

\bibitem[{{G{\"u}ver} \& {{\"O}zel}(2009)}]{Guver2009}
{G{\"u}ver}, T., \& {{\"O}zel}, F. 2009, \mnras, 400, 2050

\bibitem[{{Hasegawa} {et~al.}(1992){Hasegawa}, {Herbst}, \& {Leung}}]{hase1992}
{Hasegawa}, T.~I., {Herbst}, E., \& {Leung}, C.~M. 1992, \apjs, 82, 167

\bibitem[{{Hincelin} {et~al.}(2011){Hincelin}, {Wakelam}, {Hersant},
  {Guilloteau}, {Loison}, {Honvault}, \& {Troe}}]{Hincelin2011}
{Hincelin}, U., {Wakelam}, V., {Hersant}, F., {et~al.} 2011, \aap, 530, A61

\bibitem[{{Hocuk} {et~al.}(2017){Hocuk}, {Sz{\H{u}}cs}, {Caselli}, {Cazaux},
  {Spaans}, \& {Esplugues}}]{Hocuk2017}
{Hocuk}, S., {Sz{\H{u}}cs}, L., {Caselli}, P., {et~al.} 2017, \aap, 604, A58

\bibitem[{{Kauffmann} {et~al.}(2008){Kauffmann}, {Bertoldi}, {Bourke}, {Evans},
  \& {Lee}}]{Kauffmann2008}
{Kauffmann}, J., {Bertoldi}, F., {Bourke}, T.~L., {Evans}, II, N.~J., \& {Lee},
  C.~W. 2008, \aap, 487, 993

\bibitem[{{Kim} {et~al.}(1996){Kim}, {Kawamura}, \& {Fukui}}]{Kim1996}
{Kim}, B.~G., {Kawamura}, A., \& {Fukui}, Y. 1996, Journal of Korean
  Astronomical Society Supplement, 29, S193

\bibitem[{{Leger} {et~al.}(1985){Leger}, {Jura}, \& {Omont}}]{Leger1985}
{Leger}, A., {Jura}, M., \& {Omont}, A. 1985, \aap, 144, 147

\bibitem[{{Lippok} {et~al.}(2013){Lippok}, {Launhardt}, {Semenov}, {Stutz},
  {Balog}, {Henning}, {Krause}, {Linz}, {Nielbock}, {Pavlyuchenkov},
  {Schmalzl}, {Schmiedeke}, \& {Bieging}}]{Lippok2013}
{Lippok}, N., {Launhardt}, R., {Semenov}, D., {et~al.} 2013, \aap, 560, A41

\bibitem[{{Liseau} {et~al.}(2015){Liseau}, {Larsson}, {Lunttila}, {Olberg},
  {Rydbeck}, {Bergman}, {Justtanont}, {Olofsson}, \& {de Vries}}]{Liseau2015}
{Liseau}, R., {Larsson}, B., {Lunttila}, T., {et~al.} 2015, \aap, 578, A131

\bibitem[{{Liu} {et~al.}(2018{\natexlab{a}}){Liu}, {Stutz}, \&
  {Yuan}}]{Liuhongli2018}
{Liu}, H.-L., {Stutz}, A., \& {Yuan}, J.-H. 2018{\natexlab{a}}, \mnras, 478,
  2119

\bibitem[{{Liu} {et~al.}(2018{\natexlab{b}}){Liu}, {Kim}, {Juvela}, {Wang},
  {Tatematsu}, {Di Francesco}, {Liu}, {Wu}, {Thompson}, \&
  {Fuller}}]{Liutie2018}
{Liu}, T., {Kim}, K.-T., {Juvela}, M., {et~al.} 2018{\natexlab{b}}, \apjs, 234,
  28

\bibitem[{{Majumdar} {et~al.}(2017){Majumdar}, {Gratier}, {Ruaud}, {Wakelam},
  {Vastel}, {Sipil{\"a}}, {Hersant}, {Dutrey}, \& {Guilloteau}}]{Majumdar2017}
{Majumdar}, L., {Gratier}, P., {Ruaud}, M., {et~al.} 2017, \mnras, 466, 4470

\bibitem[{{Marsh} {et~al.}(2017){Marsh}, {Whitworth}, {Lomax}, {Ragan},
  {Becciani}, {Cambr{\'e}sy}, {Di Giorgio}, {Eden}, {Elia}, {Kacsuk},
  {Molinari}, {Palmeirim}, {Pezzuto}, {Schneider}, {Sciacca}, \&
  {Vitello}}]{Marsh2017}
{Marsh}, K.~A., {Whitworth}, A.~P., {Lomax}, O., {et~al.} 2017, \mnras, 471,
  2730

\bibitem[{{Nagy} {et~al.}(2019){Nagy}, {Spezzano}, {Caselli}, {Vasyunin},
  {Tafalla}, {Bizzocchi}, {Prudenzano}, \& {Redaelli}}]{Nagy2019}
{Nagy}, Z., {Spezzano}, S., {Caselli}, P., {et~al.} 2019, arXiv e-prints,
  arXiv:1904.01136

\bibitem[{{Newville} {et~al.}(2014){Newville}, {Stensitzki}, {Allen}, \&
  {Ingargiola}}]{LMFIT}
{Newville}, M., {Stensitzki}, T., {Allen}, D.~B., \& {Ingargiola}, A. 2014,
  {LMFIT: Non-Linear Least-Square Minimization and Curve-Fitting for Python},
  Zenodo, doi:10.5281/zenodo.11813

\bibitem[{{Ossenkopf} \& {Henning}(1994)}]{Ossenkopf1994}
{Ossenkopf}, V., \& {Henning}, T. 1994, \aap, 291, 943

\bibitem[{{Ostriker}(1964)}]{Ostriker1964}
{Ostriker}, J. 1964, \apj, 140, 1529

\bibitem[{{Planck Collaboration} {et~al.}(2011){Planck Collaboration}, {Ade},
  {Aghanim}, {Arnaud}, {Ashdown}, {Aumont}, {Baccigalupi}, {Balbi}, {Banday},
  {Barreiro}, {Bartlett}, {Battaner}, {Benabed}, {Beno{\^\i}t}, {Bernard},
  {Bersanelli}, {Bhatia}, {Bonaldi}, {Bonavera}, {Bond}, {Borrill}, {Bouchet},
  {Bucher}, {Burigana}, {Butler}, {Cabella}, {Cantalupo}, {Cappellini},
  {Cardoso}, {Carvalho}, {Catalano}, {Cay{\'o}n}, {Challinor}, {Chamballu},
  {Chary}, {Chen}, {Chiang}, {Chiang}, {Christensen}, {Clements}, {Colombi},
  {Couchot}, {Coulais}, {Crill}, {Cuttaia}, {Danese}, {Davis}, {de Bernardis},
  {de Rosa}, {de Zotti}, {Delabrouille}, {Delouis}, {D{\'e}sert}, {Dickinson},
  {Diego}, {Dolag}, {Dole}, {Donzelli}, {Dor{\'e}}, {D{\"o}rl}, {Douspis},
  {Dupac}, {Efstathiou}, {En{\ss}lin}, {Eriksen}, {Finelli}, {Forni},
  {Fosalba}, {Frailis}, {Franceschi}, {Galeotta}, {Ganga}, {Giard},
  {Giraud-H{\'e}raud}, {Gonz{\'a}lez-Nuevo}, {G{\'o}rski}, {Gratton},
  {Gregorio}, {Gruppuso}, {Haissinski}, {Hansen}, {Harrison}, {Helou},
  {Henrot-Versill{\'e}}, {Hern{\'a}ndez-Monteagudo}, {Herranz}, {Hildebrandt},
  {Hivon}, {Hobson}, {Holmes}, {Hornstrup}, {Hovest}, {Hoyland},
  {Huffenberger}, {Huynh}, {Jaffe}, {Jones}, {Juvela}, {Keih{\"a}nen},
  {Keskitalo}, {Kisner}, {Kneissl}, {Knox}, {Kurki-Suonio}, {Lagache},
  {L{\"a}hteenm{\"a}ki}, {Lamarre}, {Lasenby}, {Laureijs}, {Lawrence}, {Leach},
  {Leahy}, {Leonardi}, {Le{\'o}n-Tavares}, {Leroy}, {Lilje},
  {Linden-V{\o}rnle}, {L{\'o}pez-Caniego}, {Lubin}, {Mac{\'\i}as-P{\'e}rez},
  {MacTavish}, {Maffei}, {Maggio}, {Maino}, {Mandolesi}, {Mann}, {Maris},
  {Marleau}, {Marshall}, {Mart{\'\i}nez-Gonz{\'a}lez}, {Masi}, {Massardi},
  {Matarrese}, {Matthai}, {Mazzotta}, {McGehee}, {Meinhold}, {Melchiorri},
  {Melin}, {Mendes}, {Mennella}, {Mitra}, {Miville-Desch{\^e}nes}, {Moneti},
  {Montier}, {Morgante}, {Mortlock}, {Munshi}, {Murphy}, {Naselsky}, {Natoli},
  {Netterfield}, {N{\o}rgaard- Nielsen}, {Noviello}, {Novikov}, {Novikov},
  {O'Dwyer}, {Osborne}, {Pajot}, {Paladini}, {Partridge}, {Pasian},
  {Patanchon}, {Pearson}, {Perdereau}, {Perotto}, {Perrotta}, {Piacentini},
  {Piat}, {Piffaretti}, {Plaszczynski}, {Platania}, {Pointecouteau}, {Polenta},
  {Ponthieu}, {Poutanen}, {Pratt}, {Pr{\'e}zeau}, {Prunet}, {Puget}, {Rachen},
  {Reach}, {Rebolo}, {Reinecke}, {Renault}, {Ricciardi}, {Riller},
  {Ristorcelli}, {Rocha}, {Rosset}, {Rowan-Robinson}, {Rubi{\~n}o-Mart{\'\i}n},
  {Rusholme}, {Sajina}, {Sandri}, {Santos}, {Savini}, {Schaefer}, {Scott},
  {Seiffert}, {Shellard}, {Smoot}, {Starck}, {Stivoli}, {Stolyarov},
  {Sudiwala}, {Sunyaev}, {Sygnet}, {Tauber}, {Tavagnacco}, {Terenzi},
  {Toffolatti}, {Tomasi}, {Torre}, {Tristram}, {Tuovinen}, {T{\"u}rler},
  {Umana}, {Valenziano}, {Valiviita}, {Varis}, {Vielva}, {Villa}, {Vittorio},
  {Wade}, {Wandelt}, {White}, {Wilkinson}, {Yvon}, {Zacchei}, \&
  {Zonca}}]{Planck2011c}
{Planck Collaboration}, {Ade}, P.~A.~R., {Aghanim}, N., {et~al.} 2011, \aap,
  536, A7

\bibitem[{{Planck Collaboration} {et~al.}(2016){Planck Collaboration}, {Ade},
  {Aghanim}, {Arnaud}, {Ashdown}, {Aumont}, {Baccigalupi}, {Banday},
  {Barreiro}, {Bartolo}, \& et~al.}]{Planck2016}
---. 2016, \aap, 594, A28

\bibitem[{{Plummer}(1911)}]{Plummer1911}
{Plummer}, H.~C. 1911, \mnras, 71, 460

\bibitem[{{Redaelli} {et~al.}(2019){Redaelli}, {Bizzocchi}, {Caselli},
  {Sipil{\"a}}, {Lattanzi}, {Giuliano}, \& {Spezzano}}]{Redaelli2019}
{Redaelli}, E., {Bizzocchi}, L., {Caselli}, P., {et~al.} 2019, \aap, 629, A15

\bibitem[{{Semenov} {et~al.}(2010){Semenov}, {Hersant}, {Wakelam}, {Dutrey},
  {Chapillon}, {Guilloteau}, {Henning}, {Launhardt}, {Pi{\'e}tu}, \&
  {Schreyer}}]{seme2010}
{Semenov}, D., {Hersant}, F., {Wakelam}, V., {et~al.} 2010, \aap, 522, A42

\bibitem[{{Seo} {et~al.}(2019){Seo}, {Majumdar}, {Goldsmith}, {Shirley},
  {Willacy}, {Ward-Thompson}, {Friesen}, {Frayer}, {Church}, {Chung}, {Cleary},
  {Cunningham}, {Devaraj}, {Egan}, {Gaier}, {Gawand e}, {Gundersen}, {Harris},
  {Kangaslahti}, {Readhead}, {Samoska}, {Sieth}, {Stennes}, {Voll}, \&
  {White}}]{Seo2019}
{Seo}, Y.~M., {Majumdar}, L., {Goldsmith}, P.~F., {et~al.} 2019, \apj, 871, 134

\bibitem[{{Shinnaga} {et~al.}(2004){Shinnaga}, {Ohashi}, {Lee}, \&
  {Moriarty-Schieven}}]{Shinnaga2004}
{Shinnaga}, H., {Ohashi}, N., {Lee}, S.-W., \& {Moriarty-Schieven}, G.~H. 2004,
  \apj, 601, 962

\bibitem[{{Smith} {et~al.}(2014){Smith}, {Glover}, \& {Klessen}}]{Smith2014}
{Smith}, R.~J., {Glover}, S.~C.~O., \& {Klessen}, R.~S. 2014, \mnras, 445, 2900

\bibitem[{{Spezzano} {et~al.}(2016){Spezzano}, {Bizzocchi}, {Caselli}, {Harju},
  \& {Br{\"u}nken}}]{Spezzano2016}
{Spezzano}, S., {Bizzocchi}, L., {Caselli}, P., {Harju}, J., \& {Br{\"u}nken},
  S. 2016, \aap, 592, L11

\bibitem[{{Spezzano} {et~al.}(2017){Spezzano}, {Caselli}, {Bizzocchi},
  {Giuliano}, \& {Lattanzi}}]{Spezz2017}
{Spezzano}, S., {Caselli}, P., {Bizzocchi}, L., {Giuliano}, B.~M., \&
  {Lattanzi}, V. 2017, \aap, 606, A82

\bibitem[{{Tafalla} {et~al.}(2002){Tafalla}, {Myers}, {Caselli}, {Walmsley}, \&
  {Comito}}]{Tafalla2002}
{Tafalla}, M., {Myers}, P.~C., {Caselli}, P., {Walmsley}, C.~M., \& {Comito},
  C. 2002, \apj, 569, 815

\bibitem[{{Tang} {et~al.}(2018){Tang}, {Liu}, {Qin}, {Kim}, {Wu}, {Tatematsu},
  {Yuan}, {Wang}, {Parsons}, {Koch}, {Sanhueza}, {Ward-Thompson}, {T{\'o}th},
  {Soam}, {Lee}, {Eden}, {Di Francesco}, {Rawlings}, {Rawlings}, {Montillaud},
  {Zhang}, \& {Cunningham}}]{Tang2018}
{Tang}, M., {Liu}, T., {Qin}, S.-L., {et~al.} 2018, \apj, 856, 141

\bibitem[{{Tatematsu} {et~al.}(2017){Tatematsu}, {Liu}, {Ohashi}, {Sanhueza},
  {Nguyen Lu'o'ng}, {Hirota}, {Liu}, {Hirano}, {Choi}, {Kang}, {Thompson},
  {Fuller}, {Wu}, {Li}, {Di Francesco}, {Kim}, {Wang}, {Ristorcelli}, {Juvela},
  {Shinnaga}, {Cunningham}, {Saito}, {Lee}, {T{\'o}th}, {He}, {Sakai}, {Kim},
  {JCMT Large Program ``SCOPE'' Collaboration}, \& {TRAO Key Science Program
  ``TOP'' Collaboration}}]{Tate2017}
{Tatematsu}, K., {Liu}, T., {Ohashi}, S., {et~al.} 2017, \apjs, 228, 12

\bibitem[{{Wakelam} \& {Herbst}(2008)}]{Wakelam2008}
{Wakelam}, V., \& {Herbst}, E. 2008, \apj, 680, 371

\end{thebibliography}


\end{document}